\RequirePackage{rotating}
\PassOptionsToPackage{usenames,dvipsnames}{xcolour}
\documentclass[twocolumn,twocolappendix]{openjournal}

\pdfoutput=1 
\usepackage{amsmath,amstext}
\usepackage[T1]{fontenc}
\usepackage{apjfonts}
\usepackage{ae,aecompl}
\usepackage[utf8]{inputenc}
\usepackage{hyperref}
\usepackage[figure,figure*]{hypcap}
\usepackage{natbib}
\usepackage{url}
\usepackage{mdwlist}
\usepackage{listings}
\usepackage{rotating}
\usepackage{multirow}
\usepackage{lineno}

\shorttitle{DESC Tomography Challenge}
\shortauthors{Zuntz et al.\ (LSST~DESC)}

\begin{document}
\title{The LSST-DESC 3x2pt Tomography Optimization Challenge}
\author{Joe Zuntz\altaffilmark{1}, Fran\c{c}ois Lanusse\altaffilmark{2}, Alex I. Malz\altaffilmark{3}, Angus H. Wright\altaffilmark{3}, An\v{z}e Slosar\altaffilmark{4}, Bela Abolfathi\altaffilmark{5}, David Alonso\altaffilmark{6}, Abby Bault\altaffilmark{5}, Cl\'{e}cio R. Bom\altaffilmark{7}\altaffilmark{8}, Massimo Brescia\altaffilmark{9}, Adam Broussard\altaffilmark{10}, Jean-Eric Campagne\altaffilmark{11}, Stefano Cavuoti\altaffilmark{9}\altaffilmark{12}\altaffilmark{13}, Eduardo S. Cypriano\altaffilmark{14}, Bernardo M. O. Fraga\altaffilmark{7}, Eric Gawiser\altaffilmark{10}, Elizabeth J. Gonzalez\altaffilmark{7}\altaffilmark{15}\altaffilmark{16}, Dylan Green\altaffilmark{5}, Peter Hatfield\altaffilmark{6}, Kartheik Iyer\altaffilmark{17}, David Kirkby\altaffilmark{5}, Andrina Nicola\altaffilmark{18}, Erfan Nourbakhsh\altaffilmark{19}, Andy Park\altaffilmark{5}, Gabriel Teixeira\altaffilmark{7}, Katrin Heitmann\altaffilmark{20}, Eve Kovacs\altaffilmark{20}, Yao-Yuan Mao\altaffilmark{21},}
\collaboration{The LSST Dark Energy Science Collaboration, LSST DESC}

\footnote{Author affiliations may be found before the references.}

\begin{abstract}
This paper presents the results of the Rubin Observatory Dark Energy Science Collaboration (DESC) 3x2pt tomography challenge, which served as a first step toward optimizing the tomographic binning strategy for the main DESC analysis. 
The task of choosing an optimal tomographic binning scheme for a photometric survey is made particularly delicate in the context of a metacalibrated lensing catalogue, as only the photometry from the bands included in the metacalibration process (usually {\em riz} and potentially {\em g}) can be used in sample definition.
The goal of the challenge was to collect and compare bin assignment strategies under various metrics of a standard 3x2pt cosmology analysis in a highly idealized setting to establish a baseline for realistically complex follow-up studies; 
in this preliminary study, we used two sets of cosmological simulations of galaxy redshifts and photometry under a simple noise model neglecting photometric outliers and variation in observing conditions, and contributed algorithms were provided with a representative and complete training set.
We review and evaluate the entries to the challenge, finding that even from this limited photometry information, multiple algorithms can separate tomographic bins reasonably well, reaching figures-of-merit scores close to the attainable maximum.
We further find that adding the \emph{g} band to \emph{riz} photometry improves metric performance by $\sim15\%$ and that the optimal bin assignment strategy depends strongly on the science case: which figure-of-merit is to be optimized, and which observables (clustering, lensing, or both) are included.
\end{abstract}



\section{Introduction}
Weak gravitational lensing (WL) has emerged over the last decade as a powerful
cosmological probe \citep{cfhtlens,rcslens,desy3a,desy3b,kids,hsc}.  WL
uses measurements of coherent shear distortion to the observed shapes of galaxies
to track the evolution of large-scale gravitational fields.  It measures the integrated
gravitational potential along lines of sight to source galaxies, and can thence constrain
the laws of gravity, the expansion history of the Universe, and the history and growth
of cosmic structure.

WL has proven especially powerful in combination with galaxy clustering measurements,
which can measure the density of matter up to an unknown bias function.  The high signal
to noise of such measurements and the relative certainty of the redshift of these foreground
samples breaks degeneracies in the systematic errors that affect WL.

The \emph{3$\times$2pt} method has become a standard tool for performing this combination.
In this method, two-point correlations are computed among and between two samples, the shapes of 
background (source) galaxies and the locations of foreground (lens) galaxies, which trace foreground
dark matter haloes.  The three combinations (source-source, source-lens, and lens-lens) are
measured in either Fourier or configuration space, and can be predicted from a combination of 
perturbation theory and simulation results.  The method has been used in the Dark Energy Survey, DES, 
\citep{des-3x2pt, desy3-3x2pt}, and to combine the Kilo-Degree 
Survey, KiDS with spectroscopic surveys \citep{kids-3x2pt, kids_gama, kids_2df}.

Most lensing and 3x2pt analyses have chosen to analyze data \emph{tomographically}, 
binning galaxies by redshift.
This approach captures almost all the available information in lensing data, since lensing measures
an integrated effect and so galaxies nearby in redshift probe very similar fields.  For photometric
foreground samples, tomography also loses little information when reasonably narrow bins are used,
since redshift estimates of such galaxies have
large uncertainties\footnote{Spectroscopic foreground samples may be more likely to see significant 
gains from moving beyond tomographic methods.}.  Binning galaxies by approximate redshift also lets us 
model galaxy bias, intrinsic alignments, and other systematic errors en masse in a more tractable way.
While fully 3D methods have been proposed, prototyped, and shown to have significant promise, 
\citep{heavens,kitching}, the tomographic approach remains the standard within the field.
Tomographic 3x2pt measurements will be a key science goal in the upcoming \emph{Stage IV} surveys,
including the Vera C. Rubin Observatory \citep{rubin} and the Euclid and Roman space telescopes
\citep{euclid,roman}.

We are free to assign galaxies to different tomographic bins in any way we wish; changing the choice
can potentially affect the ease with which we can calibrate
the bins, but any choice can lead to correct cosmology results. The bins need not even be assigned contiguous redshift ranges: we can happily correlate bins with multiple redshifts if that is useful, or by some other galaxy property than redshift entirely.
We should choose, then, an assignment algorithm that maximises the constraining power for a science
case of interest.  This challenge explores such algorithms.

Various recent work have explored general tomographic binning strategies. The general question of optimization
was recently discussed in detail
in \citet{rainbow} using a self-organizing map (SOM) approach to target up to five tomographic bins.  For this
configuration they find that equally spaced redshift bins are a good proxy
for an optimized method, but note, importantly, that we should not be constrained to trying to directly
match our bin selections to specific tomographic bin edges. 
They also highlight the utility of rejecting outlier or hard-to-classify objects from samples, and the ability of SOMs
to do so cleanly.
\citet{taylor18} considered using
fine tomographic bins as an alternative to fully 3D lensing, and found that the strategy of equal numbers
per bin is less effective at high redshift, a result we will echo in this paper. The fine-binning strategy
also allows a set of nulling methods to be applied, offering various advantages in avoiding poorly-understood
regimes \citep{taylor18b,bnt,xcut}.
Most recently, \citet{euclid-nz} studied this question with respect to the Euclid mission, using the \textit{Flagship}
simulation and applying realistic photometric redshift estimation methods to the
sample (we defer the latter in this work). They highlight the power of using large numbers of bins,
differences in ideal binnings when
changing the choice of data (the inclusion galaxy galaxy-lensing), and
note the value in discarding galaxies with high redshift uncertainty. We will echo many of these issues below.

This paper is part of preparations for the analysis to be run by the Dark Energy Science Collaboration (DESC) of Legacy Survey of Space and Time (LSST) data from the Rubin Observatory.
In it, we discuss and evaluate the related challenge
of using a limited set of colour bands to make those assignments, motivated by requirements of using the
metacalibration method for galaxy shape measurement to limit biases.  
A similar question, with different motivations, was explored in \citet{jain}, who found that 3-band 
$gri$ tomography could be effective for tomographic selection, a result we will echo here in a new context: in this paper we describe the results of a 
challenge using simulated Rubin-like data with a range of methods submitted by different teams.  We explore how many tomographic bins we can effectively generate using only a subset of bands, and compare methods submitted to the challenge.

In \autoref{sec:motivation} we explain the need for methods that work on limited sets of bands in the context of upcoming lensing surveys. In \autoref{sec:design} we describe the design of the challenge and the simulated data used in it.
Section \ref{sec:results} briefly describes the entrants to the challenge, and discusses their performance (full
method details are given in appendix \ref{app:methods}).
 We conclude in \autoref{sec:conclusion}.

\section{Motivation}
\label{sec:motivation}

The methodology used to assign galaxies to bins faces special challenges when we use a particularly
effective approach for measuring galaxy shear, called \emph{metacalibration} (metacal),
which was introduced in \citet{sheldonhuff} and developed further in \citet{sheldon}.
In metacal, galaxy images are deconvolved from the point-spread function (PSF), sheared, and
reconvolved before measurement, allowing one to directly compute the response of a metric
to underlying shear and correct for it when computing two-point measurements.  This can
almost completely remove errors due to model and estimator (noise) bias from WL data, at least
when PSFs are well-measured and oversampled by the pixel grid.  The method has been successfully
applied in the DES Year 1 and Year 3 analyses \citep{des-y1-cat, des-y3-cat},
and application to early Rubin data is planned.

Furthermore, metacal provides a solution to the pernicious and general problem of
\emph{selection biases} in WL.  Measurements of galaxy shear are known to have noise-induced errors
that are
highly covariant with those on other galaxy properties, including size and, most importantly, flux and
magnitude.  It follows that a cut (or re-weighting) on galaxy magnitudes, or on any quantity derived
from  them, will induce a shear bias since galaxies will preferentially fall on either side of the cut
depending on their shear.  Within metacal, we can compute and thus account for this bias
very accurately, by performing the cut on both
sheared variants of the catalogue, and measuring the difference between the post-cut shear in the two
variants.  In DES the corrected biases were found up to around 4\%, far larger than
the requirements for the current generation of surveys \citep{des-y1-cat}.

In summary: metacal allows us to correct for significant selection biases, but only if all selections
are performed using only bands in which the PSF is measured well enough for deconvolution to be
performed\footnote{Alternative or extended methods to the metacalibration process described here, modelling the full likelihood space of deconvolved galaxy properties in a unified way might one day be feasible, but are likely to be computationally challenging.}.  For Rubin, this means using only the $r$, $i$, $z$, and perhaps $g$ bands.  In this work we
therefore study how well we can perform tomography using only this limited photometric information.  In
particular, this limitation prevents us from using the most obvious method for tomography,
and computing a complete photometric redshift probability density function (PDF) for each galaxy and assigning
a bin using the mean or peak of that PDF\footnote{This only prevents us using other photometry for \emph{selection}, not for \emph{characterization} after objects have been selected, such as computing the overall
number density $n(z)$ for a tomographic bin.}.

Simulation methods can also be used to correct for selection biases, provided that they match the
real data to high accuracy.  If we determine that limiting the bands we can use for tomography results
in a significant decrease in our overall 3x2pt signal-to-noise, outweighing the gain from the improved shape
calibration then this might suggest moving to rely more on such methods\footnote{One can account for
residual shear estimation uncertainty during parameter estimation by marginalizing over an unknown
factor.  Widening the prior on this factor decreases the overall constraining power of the analysis.}.
Constructing simulations with the required fidelity is challenging at Rubin-level accuracy and requires
careful comparison to deep field data.

\section{Challenge Design} \label{sec:design}

The DESC Tomographic Challenge opened in May 2020 and accepted entries until September 2020.

In the challenge, participants assigned simulated galaxies to tomographic bins,
using only their (g)riz bands and associated errors, and the galaxy size.  Their goal was to maximize the cosmological
information in the final split data set; this is generally achieved by separating different bins by 
redshift as much as possible.  Participants were free to optimize or simply select nominal (target) 
redshift bin edges from their training sample, but in either case had to assign galaxies into the 
different bins.

This meant there were two ways to obtain higher scores: either to improve the choice of nominal bin
edges, or to find better ways to assign objects to those bins.  This was a sub-optimal design decision:
separating the two issues would have enabled us to more easily delineate the advantages of different methods. 
This can be explored retrospectively, since we know what approach methods took, but would have been
easier in advance.  This and other mistakes in designing the challenge are described in \autoref{sec:mistakes}.

This differs from many redshift estimation questions because we are not at all concerned with estimates
of individual galaxy redshifts; instead, this challenge was desgined as a classification and bin-optimization
problem, amenable to more direct machine-learning methods.  As such we used the training / validation / testing
approach, as standard in machine-learning analyses.
The determination of the $n(z)$ of the recovered
bins (whether with per-galaxy estimates or otherwise) was not part of the challenge.

\subsection{Philosophy}

Since this was a preliminary challenge designed to explore whether it is possible \emph{in theory}
to use only the (g)riz bands for tomography, we chose to simplify several aspects of the data.  In realistic 
situations we will have access to limited sized training sets for WL photometric redshifts, since 
spectra are time-consuming to obtain.  These training sets will also be highly incomplete compared
to full catalogues, especially at faint magnitudes where spectroscopic coverage is sparse.  In this
challenge we avoided both these issues -- the training samples we provided (see \autoref{sec:data}) were comparable
in size to the testing sample, and drawn from the same population.  Given this, the challenge
represents a best-case scenario -- if no method succeeded on this easier case then more realistic
cases would probably be impossible.
These simplifications will increase scores overall, since one source of uncertainty in the calculation
is removed. We expect that they will disproportionately increase scores for large numbers of
tomographic bins, since the relative widening from individual galaxy uncertainties will be larger. Faint
galaxy samples, including those at higher redshift, will also be more affected.

Despite these simplifications, the other aspects of the challenge and process were designed to be
as directly relevant as possible to the particular WL science case we focus on.  The data set
was chosen to mimic the population, noise, and cuts we will use in real data (see \autoref{sec:data-proc}) 
and the metrics were designed to be as close to the science goals as possible, rather than a lower
level representation (see \autoref{sec:metrics}). The data set was also large enough to pose a 
reasonably realistic test of methods at the large data volume required for upcoming surveys, where
$10^9$ -- $10^{10}$ galaxies will be measured.

The challenge was open to any entrants, not just those already involved in DESC, but it was
advertised through Rubin channels, and so is perhaps best described as semi-public.
Participants developed and tested their methods locally but submitted code to be run as part of a central
combined analysis at the National Energy Research
Supercomputing Center (NERSC), where the final tests and metric evaluation were conducted.  No prize was offered apart from recognition.

\subsection{Data}
\label{sec:data}

\begin{figure}[htbp]
	\includegraphics[width=0.9\linewidth]{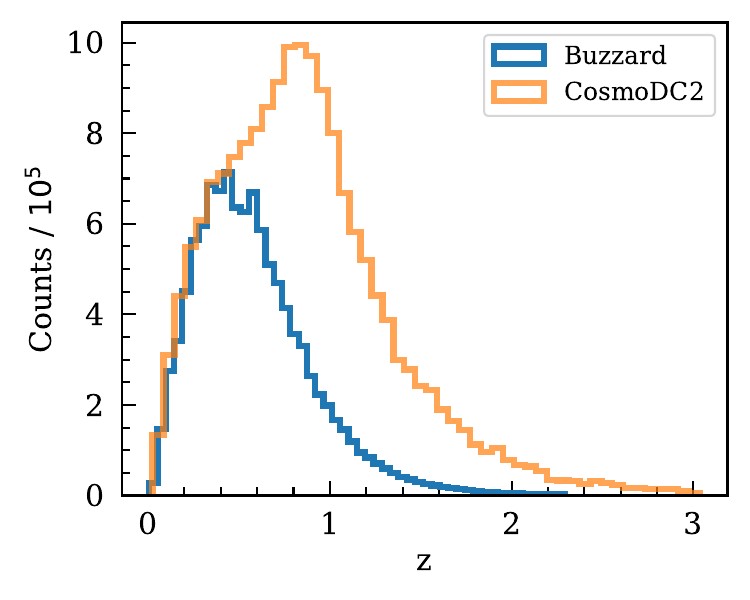}
	\caption{The underlying redshift number density $n(z)$ for the two catalogues used in the challenge,
		after the initial cuts described in the text.}
	\label{fig:initial_nz}
\end{figure}

\begin{figure}[htbp]
	\includegraphics[width=0.9\linewidth]{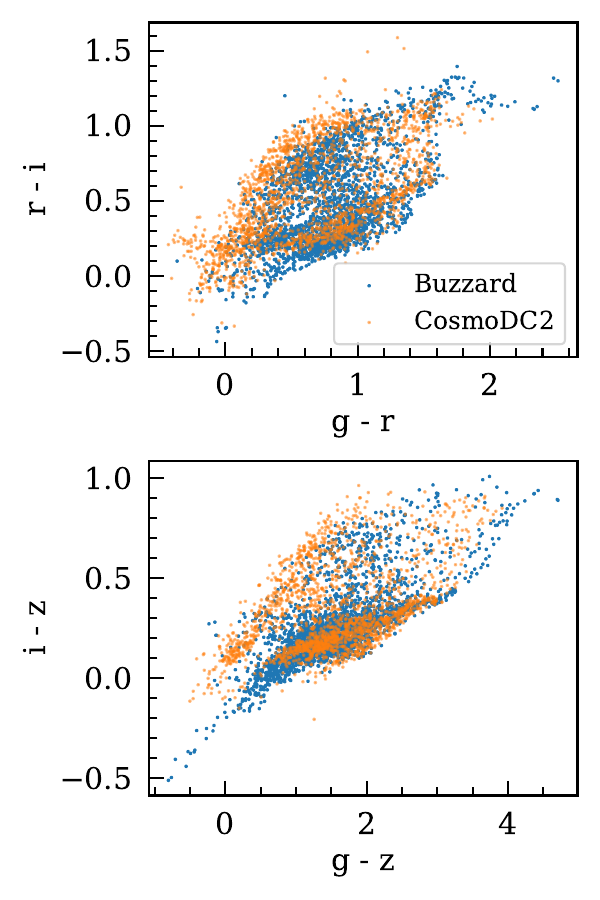}
	\caption{Colour-colour diagrams for the two catalogues used in the challenge, showing the bands
		available to participants.}
	\label{fig:colour_colour}
\end{figure}

We ran challenge entries on two data sets, CosmoDC2 and Buzzard, each of which is split into training, 
validation, and testing components.  In each case participants were supplied with the training and 
validation data, and the testing data was kept secret for final evaluation. In each case the three
data sets were random sub-selections of the full data sets, with the training and testing sets 25\% of 
the full data each and the validation 50\%. 

Using two catalogues allows us to check
for over-fitting in the design of models themselves, and thus to determine whether two
reasonable but different galaxy populations can be optimized by the same methods; this will
tell us whether our conclusions might be reasonable for real data. We discuss correlations
between metric scores on the two data sets in Section \ref{sec:metric-results}.

Each of these data sets come from cosmological simulations which provide truth values of galaxy magnitudes
and sizes.  We simulate and add noise to the objects as described below.

\subsubsection{CosmoDC2} \label{sec:cosmodc2}

The CosmoDC2 simulation was designed to support DESC and 
is described in detail in \citet{cosmodc2}.  It covers 440
deg${}^2$ of sky, and used the dark matter particles from the Outer Rim
simulations \citep{outer_rim}.  The UniverseMachine \citep{universe_machine}
simulations were then used, with the GalSampler technique \citep{galsampler},
in a combination of empirical and semi-analytic
methods, to assign galaxies with a limited set of specified properties to halos.

These objects were then matched to outputs of the Galacticus model
\citep{galacticus}, which generated complete galaxy properties and which was run
on an additional DESC simulation, AlphaQ, and included all the LSST bands.  The
simulation is complete to $r=28$, and contains objects up to redshift 3.  We
include the ultra-faint galaxies, not assigned to a specific halo, in our sample.

One limitation of the CosmoDC2 catalogues, found as part of the challenge,
was that the matching to Galacticus led to only a limited number of SEDs being
used at high redshifts, and thus too many galaxies in the simulations sharing
similar characteristics at these redshifts, such as colour-colour relations.
It was unknown whether this limitation would have any practical impact on the challenge;
this was a reason for adpoting the additional Buzzard catalogue.

\subsubsection{Buzzard}

The Buzzard catalogue was developed to support the DES Year 1 (DES-Y1)
data analysis, and is described in detail in \citet{buzzard}.
It has previously been used in DESC analyses, for example in \citet{dc1_pz}.
The catalogue used dark
matter simulations from L-GADGET2 \cite{gadget2} and then added galaxies using
an algorithm that matches a set of galaxy
property probability densities to data using sub-halo abundance matching. 
Buzzard galaxies extend to around redshift 2.3 (significantly shallower than CosmoDC2)
and were shown to be a good
match (after appropriate selections) to DES-Y1 observable catalogues.  Magnitudes
using the LSST band passes were provided in the catalogue.

\subsubsection{Post-processing} \label{sec:data-proc}

We add noise to the truth values in the extra-galactic catalogues to simulate real observations.
In each case we simulate first year Rubin observations
using the DESC {\sc TXPipe} framework\footnote{\url{https://github.com/LSSTDESC/TXPipe}}.
This follows the methodology of \citet{ivezic_jones_lupton}
and assuming the numerical values for telescope and sky properties therein:
for each galaxy it generates a Poisson sample of the number of photons per band based on
its true magnitude, the sky brightness, and instrumental characteristics.
No noise is added to the redshifts of the galaxy; we defer discussion of photometric
redshift uncertainty to future work.

In both simulations we add two cuts to approximately simulate the selection used in a real
lensing catalogue.  In both cases we apply a cut on the combined $riz$ signal to noise of
$S/N > 10$, and a size cut $T / T_\mathrm{psf} > 0.5$, where
$T$ is the trace $I_{xx} + I_{yy}$ of the moments matrix and measures the squared (deconvolved) 
radius of the
galaxy, and $T_\mathrm{psf}$ is a fixed Rubin PSF size of $0.75$ arc-seconds.

After this section, and cuts to contiguous geographic regions, we used around 20M objects from the Buzzard
simulations and around 36M objects from the CosmoDC2 simulations.  The nominal 25\% training data sets were therefore 5M and 9M objects respectively, but many entrants trained their methods on a reduced subset of the data. To make training tractable but consistent for the full suite we therefore trained all methods with 1M objects for each data set.

The overall number density $n(z)$ of the two data sets is shown in Figure \ref{fig:initial_nz},
and a set of colour-colour diagrams  in Figure \ref{fig:colour_colour}.

\subsection{Metrics}
\label{sec:metrics}

In this challenge we use only a lensing source (background) population, since the metacal requirements
do not apply to the foreground lens sample. We do, however, calculate 3x2pt metrics using the
same source and lens populations, both because this scenario is one that will be used in some analyses,
and because clustering-like statistics of lensing samples are important for studies of intrinsic
alignments. We use three different sets of metrics, one based on the overall signal to noise of the angular power spectra 
derived from the samples, and two based on a figure-of-merit for the constraints that would be obtained 
using them. The relationships between these metrics in our results are discussed in \autoref{sec:metric-results}.
We compute each metric on three different sets of summary statistics: lensing-alone, clustering-alone,
and the full 3x2pt. We therefore compute a total of nine values for each bin selection.

For each bin $i$ generated by a challenge entry, we make a histogram of the true redshifts 
of each galaxy assigned to the bin.  This is then used as our number density $n_i(z)$ in the metrics 
below. Both metrics reward bins that can
cleanly separate galaxies by redshift, since this reduces the covariance between the samples. No
method can perfectly separate them, so good methods reduce the tails of $n_i(z)$ that overlap with 
neighbouring bins.

We developed two implementations of each of these metrics, one using the Core Cosmology Library 
\citep{ccl} and one the JAX Cosmology Library \citep{jax-cosmo}, which provides differentiable theory 
predictions using the JAX library \citep{jax}.  In production we use the latter since it provided a
more stable calculation of metric 2.

Metric 1 would be approximately measurable on real data without performing a full cosmology analysis,
whereas metrics 2 and 3 would be the final output of such an analysis.  They therefore test performance at what
would be multiple stages in the wider pipeline. In each case the metrics are constructed so that a larger
value is a better score.

Entrants to the challenge were permitted to use the metrics here on the training data, and several did
so to optimize their target bin edges (see \autoref{sec:train-vs-fix}).

\subsubsection{Metric 1: SNR}

Our first metric is the total signal-to-noise ratio (SNR) of the power spectrum derived from the assigned
$n_i(z)$ values:
\begin{equation}
    S = \sqrt{\mu^{T} C^{-1} \mu}
\label{eq:snr}
\end{equation}
where $\mu$ is a stack of the theoretical predictions for the $C_\ell$ spectra for each tomographic 
bin pair, in turn (including auto- and cross-correlations for both clustering and lensing), and $C$ a estimate of the covariance between them, making the approximation that the underlying density field has a Gaussian distribution as described in
\citet{takada_jain} and summarized in \autoref{app:theory}.   We compute this metric for lensing alone, clustering alone, and the full 3x2pt combination.

\subsubsection{Metrics 2 \& 3: $w_{0}-w_{a}$ and $\Omega_c-\sigma_8$ FOMs}

Metric 2 uses a figure of merit (FOM) based on that presented by the Dark Energy Task Force (DETF) \citep{detf}.  We 
use the inverse area of a Fisher Matrix ellipse, which approximates the size in two axes 
of the posterior PDF one would obtain on analysing the spectra:
\begin{align}
    F &= \left( \frac{\partial \mu}{\partial \theta} \right)^T C^{-1} \left( \frac{\partial \mu}{\partial \theta} \right), \\
    \mathrm{FOM} &= \frac{1}{2 \pi \sqrt{\det{([F^{-1}]_{p_1, p_2})}}}
\label{eq:fom}
\end{align}
where $[F^{-1}]_{p_1, p_2}$ extracts the 2x2 matrix for two parameters.  We use both the original
DETF variant in which $p_1 = w_0$ and $p_2 = w_a$, representing the constraining power on dark energy 
evolution, and a version in which $p_1 = \Omega_c$ and $p_2 = \sigma_8$, representing constraining
power on overall cosmic structure amplitude ($\sigma_8$), and its evolution (a function of $\Omega_c$). In each case the complete parameter space consists
of the seven $w_0w_aCDM$ cosmological parameters; we used the parameter combination used natively by the DESC core cosmology library \citep{ccl} ($\Omega_c, \Omega_b, \mathrm{H}_0, \sigma_8, n_s, w_0, w_a$). This simplified measure does not include any nuisance
parameters modelling systematic errors in the analysis, most importantly the galaxy bias parameters in the clustering 
spectra; this artifically increases the constraining power of the galaxy clustering-only metric, as we will see below, but the metric still acts as a useful proxy for science cases dependent on finding well separated bins.
Like all Fisher analyses, this metric approximates posterior constraints as Gaussian. 

The JAX library provides automatic differentiation of the calculation of the $C_\ell$ values, and so avoids the painful numerical sensitivity which usually affects Fisher matrix calculations. We verified this by comparing the code to tuned finite difference estimates on CCL spectrum calculations.

\subsection{Infrastructure}

The challenge was structured as a python library, in which each assignment
method was a subclass of an abstract base parent superclass, {\sc Tomographer}.
The superclass, and other challenge machinery, performed initialization,
configuration option handling, data loading, and computing derived quantities
such as colours from magnitudes.

Participants were expected to subclass
{\sc Tomographer} and implement two methods, {\sc train} and {\sc apply}.  Each
was required to accept as an argument an array or dictionary of galaxy properties
(magnitudes, and, optionally, sizes, errors, and colours).  The {\sc train} method
was also passed an array of true redshift values, which could be used however they wished.
Methods were submitted to the challenge in the form of a pull request to the challenge
repostitory on GitHub\footnote{\url{http://github.com/LSSTDESC/tomo_challenge}}.

The training and validation parts of the data set were made available for training
and hyper-parameter tuning. Once the challenge period was complete, the algorithms
were collected from the pull requests and run at NERSC. If a method failed due to a difference in the
runtime environment or the hardware at NERSC, the participants and organizers tried
to amend it after the formal challenge period ended, though this was not always possible.

\subsection{Control Methods}

Two ``control'' methods were used to test the challenge machinery and ensure
that metrics behaved appropriately in specific limits.

The \textsc{Random} algorithm randomly assigned galaxies to each tomographic bin, resulting
in bins which each had the same $n(z)$, on average (the random noise fluctuations
are very small when using this number of galaxies).  Using this method, we expect
that increasing the number of bins should not increase any metric score, since bins are
maximally correlated.  We find this to be true for all our metrics.

The \textsc{IBandOnly} method used only the $i$-band to select bin assignments.
Edge values of bins in $i$ were chosen such that each bin had the same number of training galaxies in,
and then test sample galaxies assigned to a bin based only on their $i$ band value. The band correlates
only weakly with redshift, so the scores of the method are poor.
We expect only a small
increase in metric score with the number of bins, and also that the addition of the $g$-band
should make no difference to scores, because that column of data is not used in the method.  Again, we find this to be true.

\section{Methods \& Results} \label{sec:results}

Twenty-five methods were submitted to the challenge, including the control methods described above.
Most used machine learning methods of various types to perform the primary classification,
rather than trying to perform a full photometric redshift analysis.  
These methods are listed in \autoref{tab:entrants} and described in full in \autoref{app:methods}.

\begin{table*}[]
\centering
	\begin{tabular}{|l|p{8cm}|c|}
	\hline
	\textbf{Name} & \textbf{Description} & \textbf{Submitters} \\
	\hline
	\textsc{Random} & Random assignment (control method)  & Organizers \\
	\textsc{IBandOnly} & i-band only (control method)  & Organizers \\
	\textsc{RandomForest} & A collection of individual decision trees & Organizers \\
	\textsc{LSTM} & Long Short-Term Memory neural network & CRB, BF, GT, ESC, EJG \\
	\textsc{AutokerasLSTM} & Automaticaly configured LSTM variant. & CRB, BF, GT, ESC, EJG \\
	\textsc{LGBM} & Light Gradient Boosted Machine & CRB, BF, GT, ESC, EJG \\
	\textsc{TCN} & Temporal Convolution Network & CRB, BF, GT, ESC, EJG \\
	\textsc{CNN} & Convolutional Neural Network & CRB, BF, GT, ESC, EJG \\
	\textsc{Ensemble} & Combination of three other network methods & CRB, BF, GT, ESC, EJG \\
	\textsc{SimpleSOM} & Self-Organizing Map & AHW \\
	\textsc{PQNLD} & Extension of {\sc SimpleSom} combining with template-based redshift estimation & AHW \\
	\textsc{UTOPIA} & Nearest-neighbours, optimised in the limit of large representative training sets & AHW \\
	\textsc{ComplexSOM} & Extension of {\sc SimpleSom} with an additional assignment optimization & DA, AHW, AN, EG, BA, ABr \\
	\textsc{GPzBinning} & Gaussian Process redshift estimator and binner & PH \\
	\textsc{JaxCNN/JaxResnet} & Two related CNN-based bin edge optimizers & AP \\
	\textsc{NeuralNetwork1/2} & Dense network optimizing bin assignment for two metrics & FL \\
	\textsc{PCACluster} & Principal Component Analysis of fluxes then optimized clustering & DG \\
	\textsc{ZotBin/ZotNet} & Two related neural network methods with a preprocesing normalizing flow & DK \\
	\textsc{FFNN} & Feed-forward neural network & EN \\
	\textsc{FunBins} & Random forest with various nominal bin edge selection & ABa \\
	\textsc{MLPQNA} & Multi-layer perceptron & SC, MB \\
	\textsc{Stacked Generalization} & Combination of Gradient Boosted Trees classifiers with standard Random Forests (50/50) & JEC\\
	\hline
	\end{tabular}
	\caption{Methods entered into the challenge. The algorithms are described more fully in \autoref{app:methods}.}
	\label{tab:entrants}
\end{table*}

\subsection{Result types}

\begin{table*}[]
\centering
	\begin{tabular}{|l|llll|llll|}
		\hline
		& \multicolumn{4}{c|}{\textbf{riz}}      & \multicolumn{4}{c|}{\textbf{griz}}                                \\ \hline
		\textbf{Method} & \textbf{3-bin} & \textbf{5-bin} & \textbf{7-bin} & \textbf{9-bin} & \textbf{3-bin} & \textbf{5-bin} & \textbf{7-bin} & \textbf{9-bin} \\ \hline
		{\sc ComplexSOM } & 38.0 & 52.1    & 94.4    & 101.6    & 34.9             & 45.0             & 91.6             & 100.3\\
{\sc JaxCNN } & 59.9 & 101.9    & 105.2    & --    & 79.7             & 125.2             & 150.0             & --\\
{\sc JaxResNet } & 73.7 & 111.1    & 131.6    & --    & 82.2             & 126.0             & --             & 161.5\\
{\sc NeuralNetwork1 } & 76.3 & 117.6    & 135.8    & 109.9    & 81.9             & 132.9             & 158.0             & 96.9\\
{\sc NeuralNetwork2 } & 30.5 & 48.0    & 57.8    & 136.1    & 49.4             & 102.7             & 122.2             & 140.0\\
{\sc PCACluster } & 29.5 & 68.1    & 75.8    & *    & 50.1             & 72.7             & 90.8             & *\\
{\sc ZotBin } & 64.8 & 106.4    & 121.8    & 135.3    & 77.6             & 120.0             & 141.8             & 154.7\\
{\sc ZotNet } & 73.6 & 111.2    & 131.8    & 145.9    & 83.7             & 128.5             & 150.1             & 167.2\\
\hline
{\sc AutokerasLSTM } & 31.1 & 74.9    & 103.4    & --    & 44.1             & 67.5             & 122.2             & 98.4\\
{\sc CNN } & 27.1 & 50.8    & 76.7    & 103.3    & 30.5             & 57.7             & 95.4             & 122.4\\
{\sc ENSEMBLE } & 65.6 & 65.6    & 94.8    & *    & *             & *             & *             & *\\
{\sc FunBins } & 36.8 & 82.5    & 122.1    & 141.8    & 42.1             & 100.4             & 142.5             & 167.2\\
{\sc GPzBinning } & 26.2 & 49.7    & 80.8    & 111.7    & 27.8             & 55.3             & 87.2             & 126.9\\
{\sc IBandOnly } & 38.0 & 50.0    & 54.4    & 57.3    & 38.0             & 50.0             & 54.4             & 57.3\\
{\sc LGBM } & 27.0 & 50.8    & 78.8    & 108.2    & 30.3             & 57.4             & 92.9             & 125.2\\
{\sc LSTM } & 26.7 & 51.1    & 81.4    & 103.6    & 30.2             & 57.1             & 95.9             & 126.9\\
{\sc MLPQNA } & 39.3 & 64.7    & 93.2    & 121.8    & 42.8             & 72.2             & 109.0             & 133.7\\
{\sc Stacked Generalization } & 35.3 & 60.5    & 83.5    & 120.4    & 39.4             & 63.9             & 93.1             & 148.9\\
{\sc PQNLD } & 39.5 & 60.5    & 77.7    & 105.4    & 42.9             & 71.0             & 105.9             & 133.8\\
{\sc Random } & 1.2 & 1.2    & 1.2    & 1.2    & 1.2             & 1.2             & 1.2             & 1.2\\
{\sc RandomForest } & 39.5 & 65.2    & 93.2    & 118.5    & 43.0             & 72.8             & 110.1             & 136.5\\
{\sc SimpleSOM } & 39.5 & 59.9    & 77.9    & 98.5    & 42.2             & 73.2             & 108.0             & 134.8\\
{\sc FFNN } & 39.0 & 64.3    & 90.3    & 114.7    & 43.0             & 72.0             & 112.3             & 137.2\\
{\sc TCN } & 40.0 & 65.3    & 90.9    & 110.5    & 44.1             & 72.8             & 108.0             & 134.0\\
{\sc UTOPIA } & 39.0 & 63.7    & 90.8    & 112.2    & 43.1             & 74.0             & 112.8             & 139.3\\

		\hline
	\end{tabular}
	\caption{Values of the 3x2pt DETF ($w_0,w_a$) figure-of-merit achieved by entrant methods on the 
		CosmoDC2 part of the challenge. The horizontal line separates two general approaches to the problem, as discussed in \autoref{sec:train-vs-fix}: those below the line used fixed
		target edges for tomographic bins and optimized the assignment of objects to those bins,
		whereas those above also optimized the target edges themselves.
        Asterisks indicate runs which failed during classification, and dashes indicate runs with
        number counts too small for spectra to be generated.
        }
	\label{tab:cosmodc2}
\end{table*}

The methods described above were run on a data set, previously unseen by entrants but drawn
from the same underlying population, of size 8.6M objects for CosmoDC2 (our fiducial simulation) and 5.4M objects for Buzzard.
A complete suite of the metrics described in Section \ref{sec:metrics} was run on each of the two
for each method.  In total, each method was run on 16 scenarios, each combination of: $griz$ vs. $riz$,
Buzzard vs. CosmoDC2, and 3, 5, 7, and 9 bins requested.  Nine metrics were calculated for each
scenario: clustering, lensing, and 3x2pt for each of the SNR metric (Equation \ref{eq:snr}), the $w_0-w_a$ 
DETF FOM, and the $\Omega_c - \sigma_8$ FOM (Equation \ref{eq:fom}).

\subsection{Results Overview}
The DETF results for each method are shown in Table \ref{tab:cosmodc2}.  Complete results for all metrics
for the nine-bin scenario only are shown in appendix Tables \ref{tab:full_dc2} and \autoref{tab:full_buzz}.

In all these tables, missing entries are shown for two reasons.
The first, marked with an asterisk, is when a method did not run for the given scenario,
either due to memory overflow or taking an extremely long time.
The second, marked with a dash, is when a method ran, but generated at least one bin with an extremely 
small number count, such that the spectrum could not be calculated.

\begin{figure}[htbp]
	\includegraphics[width=1\linewidth]{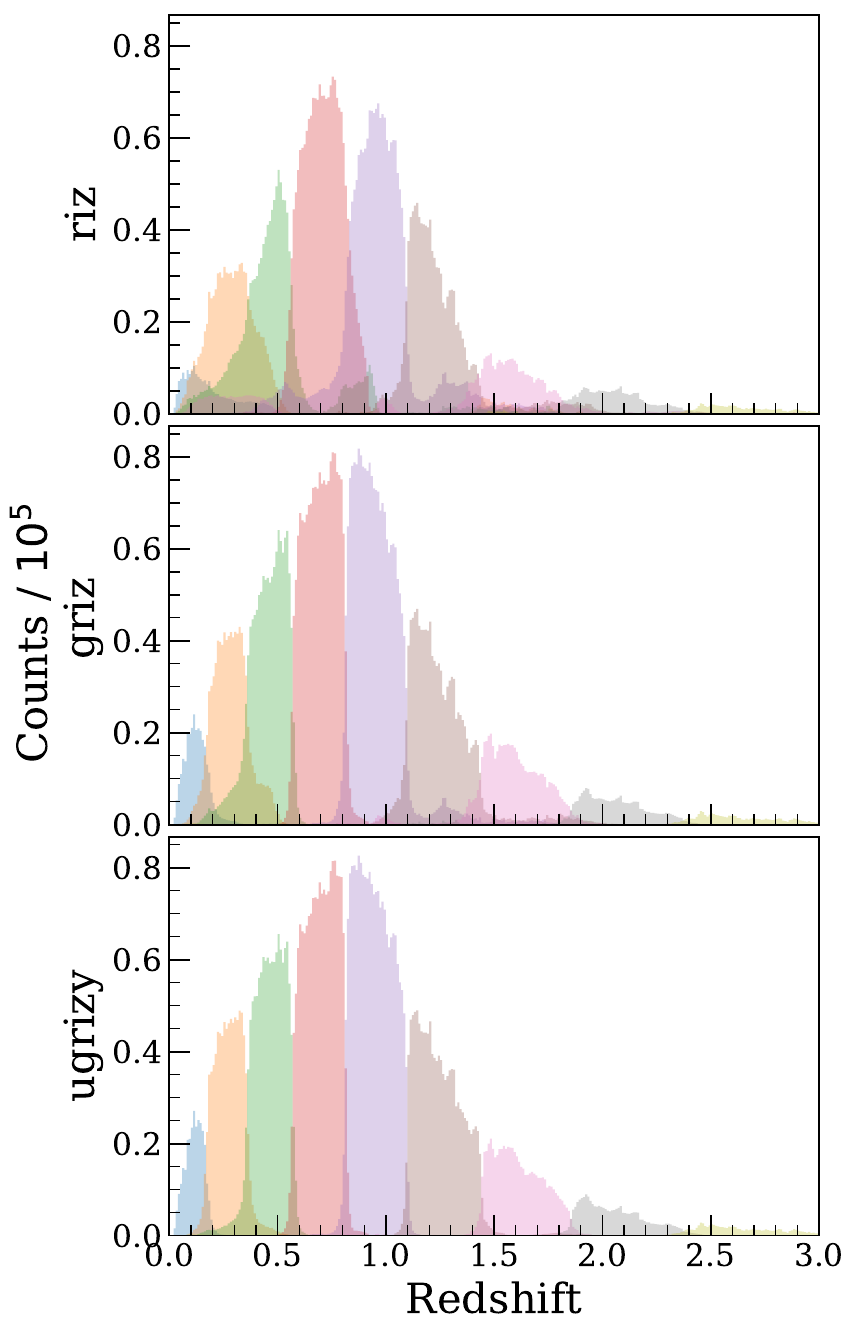}
	\caption{The tomographic $n(z)$ values generated by the winning method in the challenge, {\sc FunBins},
		generating nine bins with the CosmoDC2 data.
		The upper panel shows results for the method applied to $riz$ bands, and the middle on $griz$.
		The lower panel is for comparison and is not part of the challenge;
		it illustrates {\sc FunBins} results using all the $ugrizy$ bands.
		The increased spread and overlap of the bins as bands are removed decreases the overall constraining
		power and signal-to-noise.
	}
	\label{fig:funbin_nz}
\end{figure}

No one method dominates the metrics across the different scenarios considered.
Before the challenge began we somewhat arbitrarily chose 
the 9-bin 3x2pt $griz$ CosmoDC2 $w_0-w_a$ metric as the fiducial scenario on which
an overall winning method would be selected.
By this criterion the best method was, by a small margin, \textsc{FunBins},
which used a random forest to assign galaxies to after choosing nominal bin edges that split 
the range spanned
by the sample into equal co-moving distance bins. 
As described in \autoref{sec:funbins},  this splitting was designed
to minimize the shot noise for the angular power in each bin, and this has proven successful. 
Figure \ref{fig:funbin_nz} shows the number density obtained by this method, for both the riz and griz 
scenarios, and an additional post-challenge run using the full Rubin $ugrizy$ bands.
The metrics \textsc{FunBins}
obtained on $ugrizy$ are barely higher than those with $griz$ -- the largest
increase is 8\% for the $\Omega_c-\sigma_8$ metric, with the remainder all 5\% or less.

\begin{figure*}[htbp]
	\includegraphics[width=\linewidth]{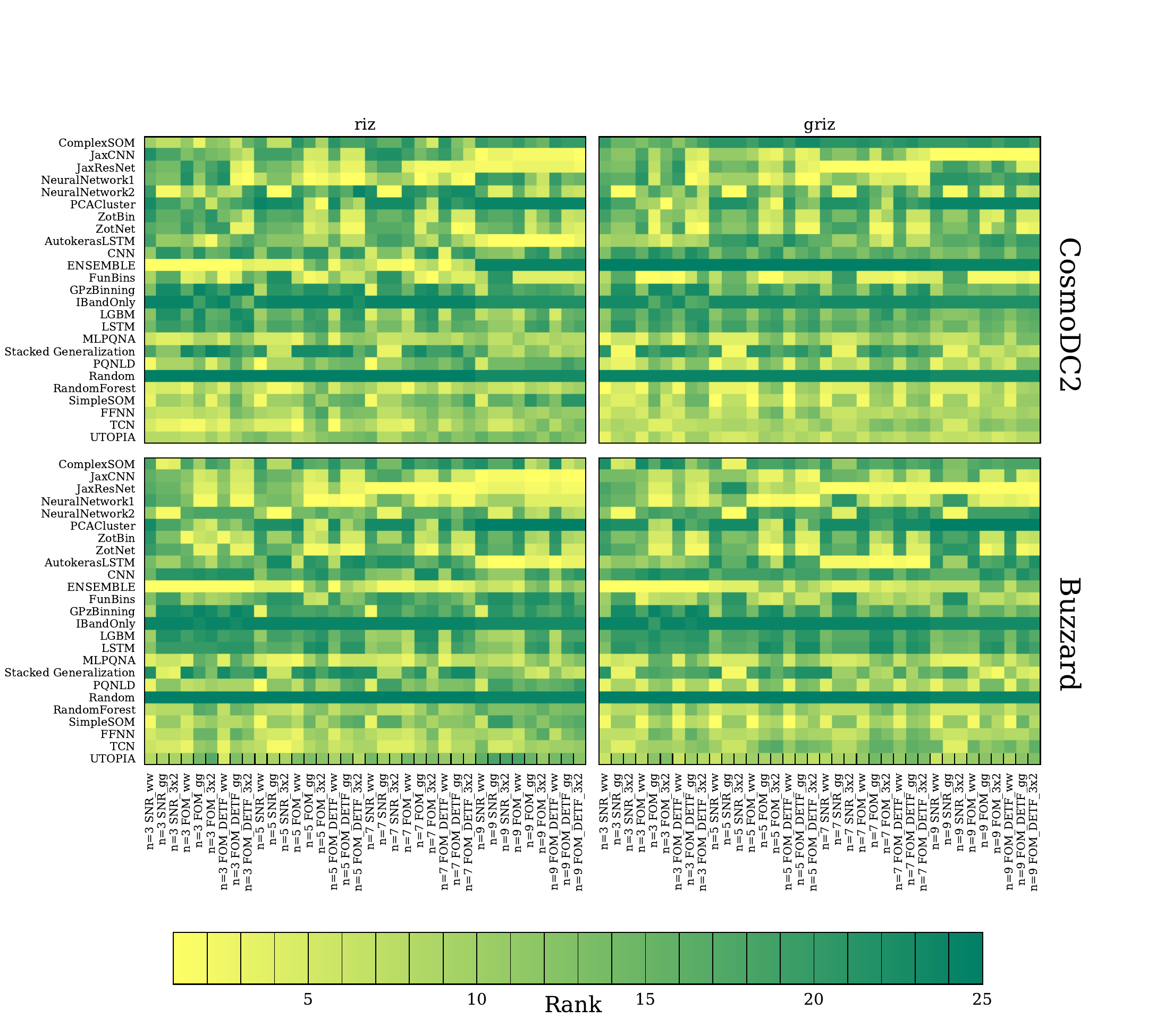}
	\caption{The rank of each method (1 = highest scoring, 25 = lowest scoring) within each configuration of the challenge, for each metric separately.  A method that was good everywhere would appear as yellow strip across an entire horizontal range.}
	\label{fig:rank_grid}
\end{figure*}

Other methods performed better in other branches of the challenge; in particular we note \textsc{ZotNet}, \textsc{Stacked Generalization}, and \textsc{NeuralNetwork} (see sections \ref{sec:zot}, \ref{sec:stackgen}, and \ref{sec:nn} respectively), which each won more than three 9-bin scenarios. An illustration of the rank achieved by each method within each challenge configuration and for each metric
is shown in \autoref{fig:rank_grid}.

These results suggest that, at least in our idealised scenario of perfect training sets and given the limitations of our simulations, restricted photometry does not necessarily catastrophically limit our ability to use tomography.
In both the CosmoDC2 and Buzzard data sets,
even the $riz$ bands alone are sufficient to split objects into nine reasonably separated
tomographic bins.  Once bins become this narrow the primary analysis concern will become the calibration 
of overall bin $n(z)$ values, rather than initial tomography, so we consider this to be reassuring.

There is a clear widening of the $n(z)$ when losing the $g$-band.  This widening leads to a loss in 
constraining power, which is discussed more in Section \ref{sec:gband}.

\begin{figure*}[htbp]
\includegraphics[width=0.9\linewidth]{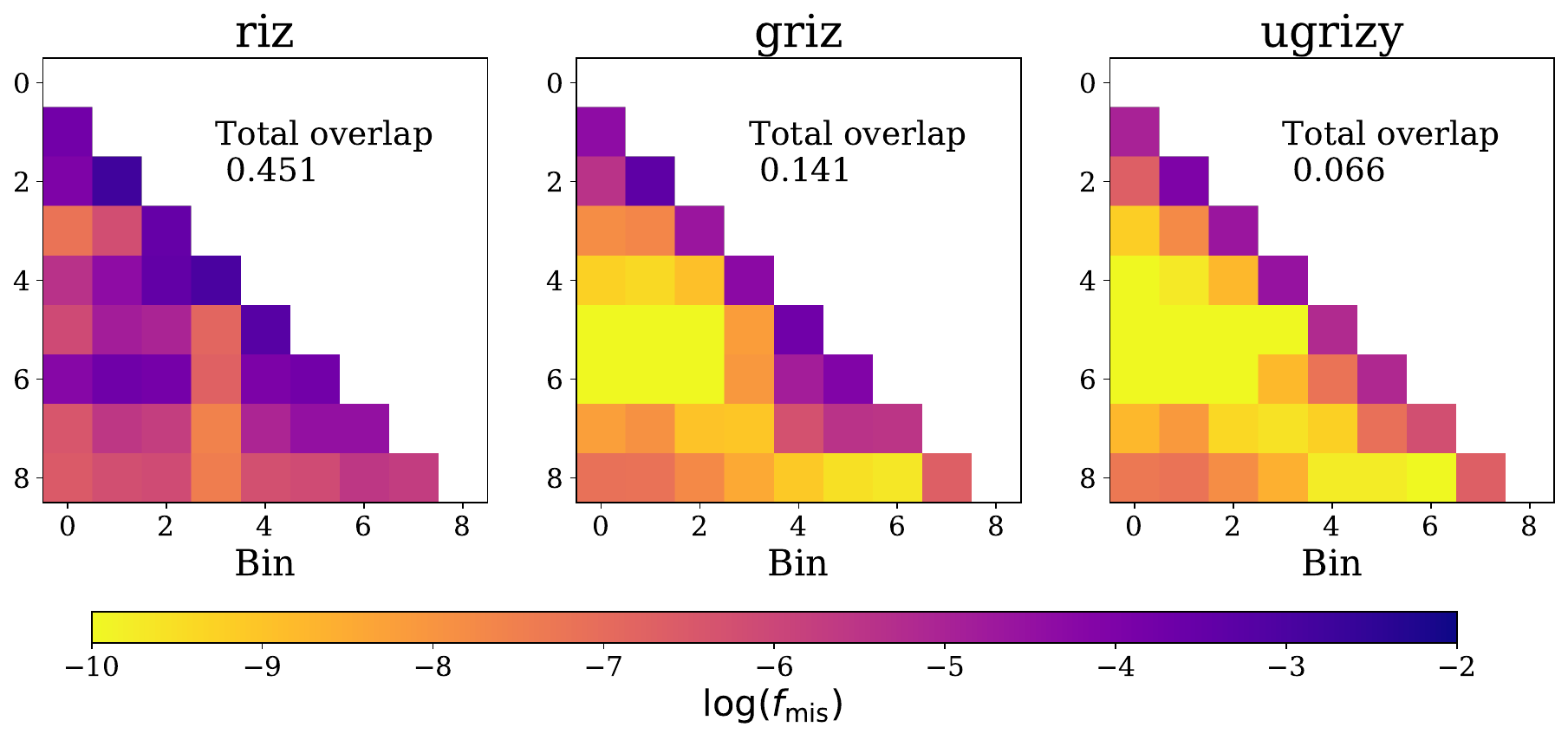}
\caption{The pairwise fractions $f_\mathrm{mis}$ of total objects in redshift regions where multiple tomographic bins contain galaxies, for 
the \textsc{FunBins} tomography method and three choices of bands.  
The $x$- and $y$-axes 
show tomographic bin indices, 
and the colour represents the fraction of objects (compared to the total number in the challenge) that are 
at redshifts where both tomographic bins have members, in bins of width $\delta z=10^{-3}$.  
This metric corresponds to
the fraction of the area that is shaded in two colours in Figure \ref{fig:funbin_nz}.
The total-overlap figure gives the total fraction of objects
that are ambiguous over all bins, and is the sum of all the pairwise values.}
\label{fig:overlap}
\end{figure*}

For comparison, we also show results using all six LSST bands, $ugrizy$; these could not be used
with the metacalibration method. The improvement in the sharpness of the bins is noticeable but not large.
We quantify this in Figure \ref{fig:overlap}, which shows the fraction of objects that are in ambiguous
redshift slices, defined as objects in redshift ranges where there are more objects assigned to another 
tomographic bin.  The total fraction of such objects is also shown. This illustrates how adding the $g$ band 
improves the assigment
for almost every pair of bins, and reduce the ambiguous fraction by a factor of three to the 
point where a significant majority of galaxies are clearly assigned.  Further adding the remaining $u$ and $y$
bands improves the assignment in a way that is smaller, though not insignificant, in both relative and absolute terms.
This latter addition also does not translate into large increases in the figures of merit: the 
DETF score for CosmoDC2 is 169.7, barely higher than its score of 167.2 for $griz$, and this is also true for other 
metrics.   While this is only a general indication and not a thorough exploration, it is encouraging.

\textsc{FunBins} seems to be a good and stable general method, though several other methods perform as well
or slightly better in other scenarios.  Notably, many methods reach the same plateau in lensing-only metrics,
suggesting this is somewhat easier to optimize; this is consistent with the general reason that photometric
redshifts suffice for lensing: the broad lensing kernels are less sensitive to small changes.

Several of the neural network methods perform well up to seven tomographic bins but fail when run with 
nine.  This is presumably not an intrinsic feature of these methods, but depends on implementation details.
Further development solving these challenges should be valuable, since they are among the highest scoring seven-bin
methods.

\subsection{Non-assignment}
The challenge did not require entrants to assign all galaxies to bins; aside from the impact of reduced final sample size on the metrics, no additional penalty was applied for discarding galaxies entirely.  Several methods
took advantage of this.  The \textsc{ComplexSOM} method (see \autoref{sec:csom}) found
it did not improve scores, but as noted in \autoref{sec:pqnld} this may be an artifact
of the high density of the training data, and so could be explored in future challenges\footnote{In real data or more realistic simulations discarding galaxies has more complicated effects than simply decreasing sample size as it does here - removing objects with broad $p(z)$ can narrow the final $n(z)$, not just lower it. This was not a factor in this challenge because we use true redshifts to calculate our metrics.}. See \citet{euclid-nz} for a discussion of this topic in the Euclid context.

\subsection{Metric comparisons} \label{sec:metric-results}
\begin{figure}[htbp]
\includegraphics[width=0.9\linewidth]{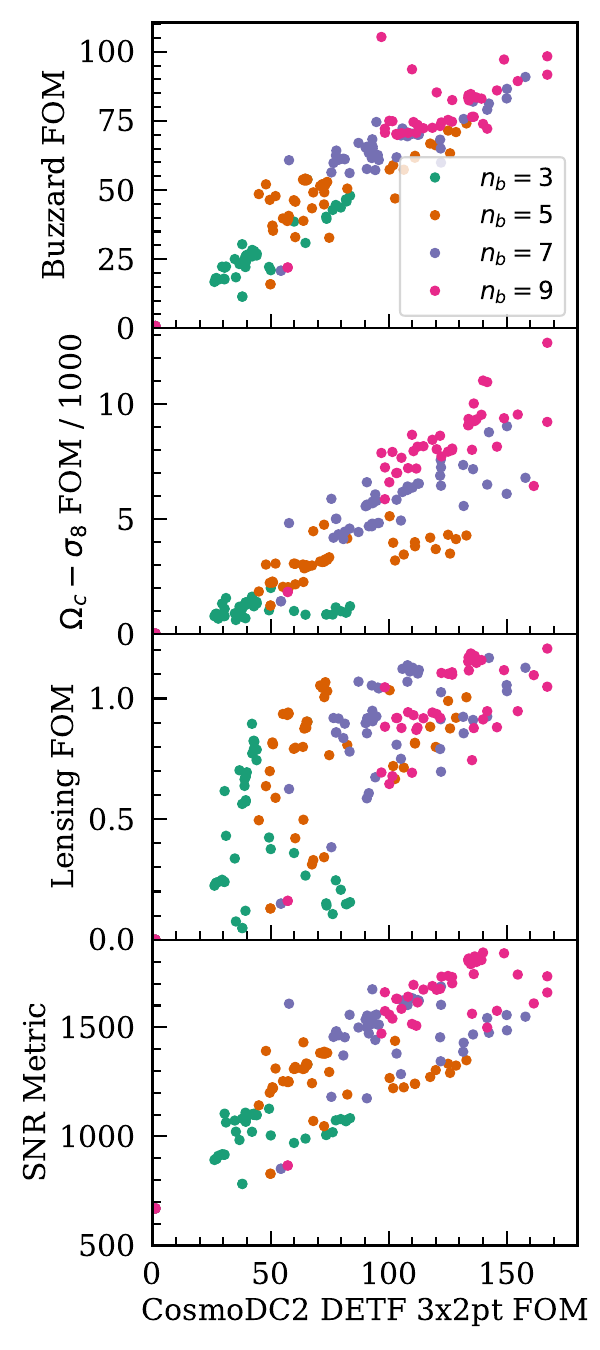}
\caption{A comparison of the metrics used in our challenge for the different methods that were entered. The x-axis is shared, and shows
the 3x2pt $w_0 - w_a$ (DETF) figure of merit on the CosmoDC2 data set for all methods, 
with colours labelling the number
of bins used.  Each of the panels shows a metric which is the same except for a single change, 
which is labelled on its y-axis.  For example, the top panel still shows the 3x2pt $w_0 - w_a$ (DETF) figure of merit,
but for the Buzzard data set.}
\label{fig:metrics}
\end{figure}

Figure \ref{fig:metrics} shows a comparison of some of the different metrics used in the challenge.
We plot, using assignments for methods and for all bin counts, the relationship between our fiducial
metric, the DETF 3x2pt on CosmoDC2, and other metrics we measured.

The different metrics are generally well-correlated (this also holds true for the metrics not 
shown here), especially at the highest values of the metrics representing the most successful
methods. This is particularly so when comparing the Buzzard and CosmoDC2 metrics, which is encouraging 
as it implies these methods are broadly stable to reasonable changes in the underlying galaxy
distribution, and hence on the survey properties.  The largest outlier comes from the 
{\sc NeuralNetwork} entry.

This is not true of the overall winning method for large bin counts - as shown in the tables
in \autoref{app:tables}, significantly more methods achieve a top-scoring spot in the Buzzard
table \ref{tab:full_buzz} than in the CosmoDC2 table \ref{tab:full_dc2}. This reinforces the
conclusion discusssed below in \autoref{sec:train-vs-fix} that target bin edges should be 
re-optimized for each new scenario, though the relatively noisy
results make this difficult to interpret.

The relatively broader spread in the lensing-only metric illustrates how our metric is dominated
by the stronger constraining power in the clustering part of the data set.  This arises because
we do not include a galaxy bias model in our FOM, and so the constraining power of the clustering
is artifically high; this limitation is one reason we explore a wide variety of metrics.

Finally, there is a bimodal structure in the relationship between the FOM and SNR metrics.
This largely traces the split between models which train the target bin edges vs those
which fix them, as described below in Section \ref{sec:train-vs-fix}.

\subsection{Impact of losing g-band} \label{sec:gband}
\begin{figure*}
\includegraphics[width=0.9\linewidth]{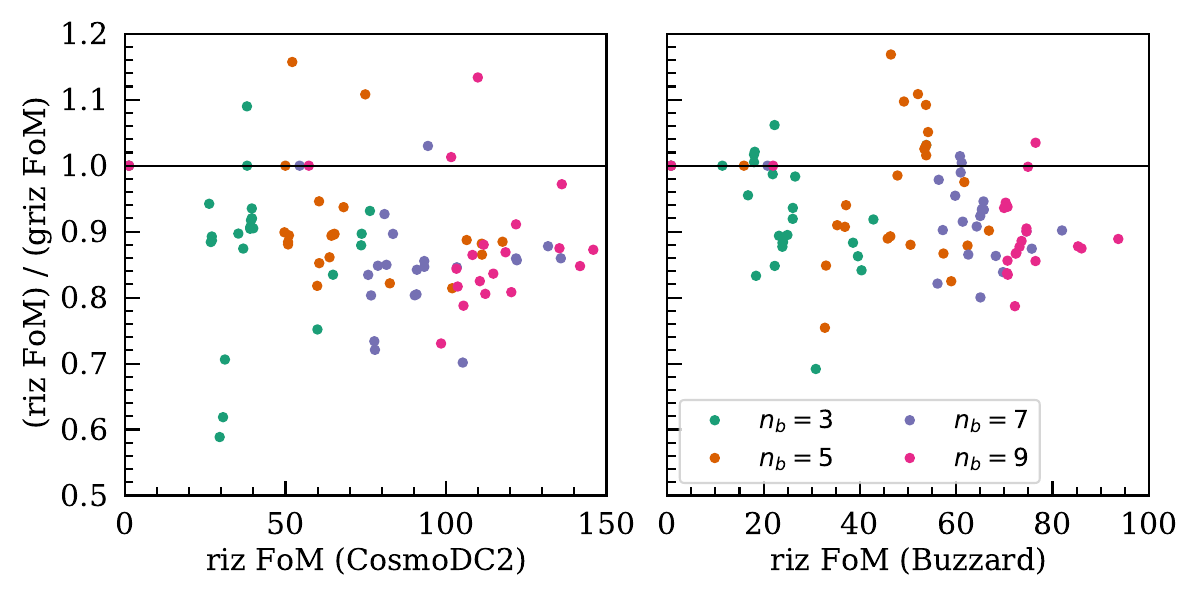}
\caption{The degradation in constraining power over all methods
 when comparing them on methods including and excluding
the g-band, in terms of the ratio of the $w_0-w_a$ figure of merit.  The left-hand panel shows
results for the CosmoDC2 scenarios, and the right-hand panel for the Buzzard scenarios.  Colours
indicate the number of bins used in the analysis. Some methods perform worse when adding the $g$-band 
and have ratios greater than unity.}
\label{fig:loss}
\end{figure*}

Figure \ref{fig:loss} shows the ratio of the FOM between methods using the griz and the same method
using the riz.  Each point represents one method with a specific number of bins, indicated by colour.
For lower scoring methods and for smaller numbers of bins there is more scatter in the ratio,
but for the highest scoring algorithms and configurations the loss when losing g-band is a little
more stable, at around 10--15\%.

The value of the extra FoM that could be gained by adding the band should be weighed against 
the challenge of high-accuracy
PSF measurement for this band, and the cost in time and effort of determining it, 
but until the end of the survey this is unlikely to be the limiting factor in overall precision.

Some methods perform worse when adding the $g$-band, which we ascribe to reduced performance
in their machine learning algorithms when increasing the dimensionality of their inputs; further tuning
of their hyperparameters could perhaps alleviate the issue.

\subsection{Trained edges vs fixed edges}\label{sec:train-vs-fix}

Some of the methods in the challenge accepted target bin edges as an input to their algorithms,
and then used classifiers to assign objects to those bins.  Entrants could select these edges; many followed the example random forest entry and split into equal counts per bin; some like \textsc{FunBins} used alternative
criteria. Other methods optimized the values
of bin edges themselves, by maximizing a metric.  For the runs shown here, these methods mostly
targeted one of the 3x2pt metrics.

\begin{figure}
\includegraphics[width=0.9\columnwidth]{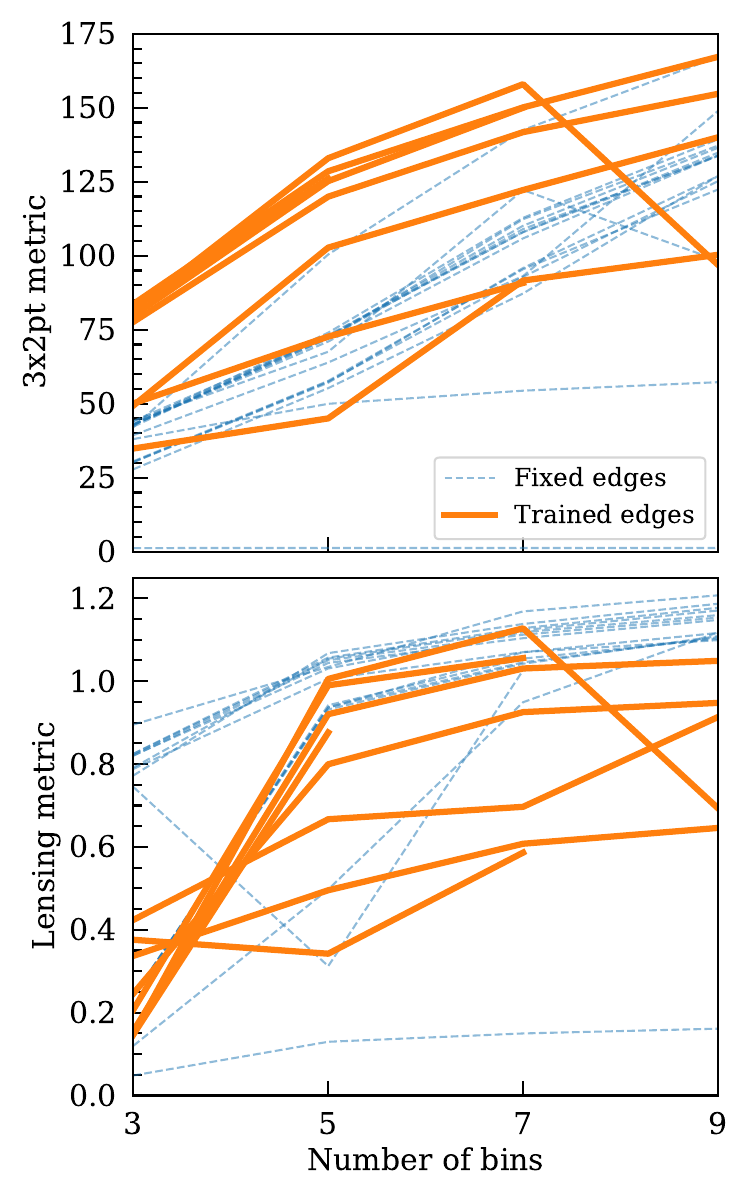}
\caption{A comparison of results from methods which used fixed target edges (dashed blue)
with those which optimized target bins (solid orange).  Methods were trained on the 3x2pt metric
shown in the upper panel; the lower performance for these classifications 
for the lensing-only metric shown in the lower panel illustrates the need for science case-specific
binning choices. In each case, classification is done using griz bands on the CosmoDC2 data set.}
\label{fig:edges}
\end{figure}

\begin{figure*}
\includegraphics[width=1\linewidth]{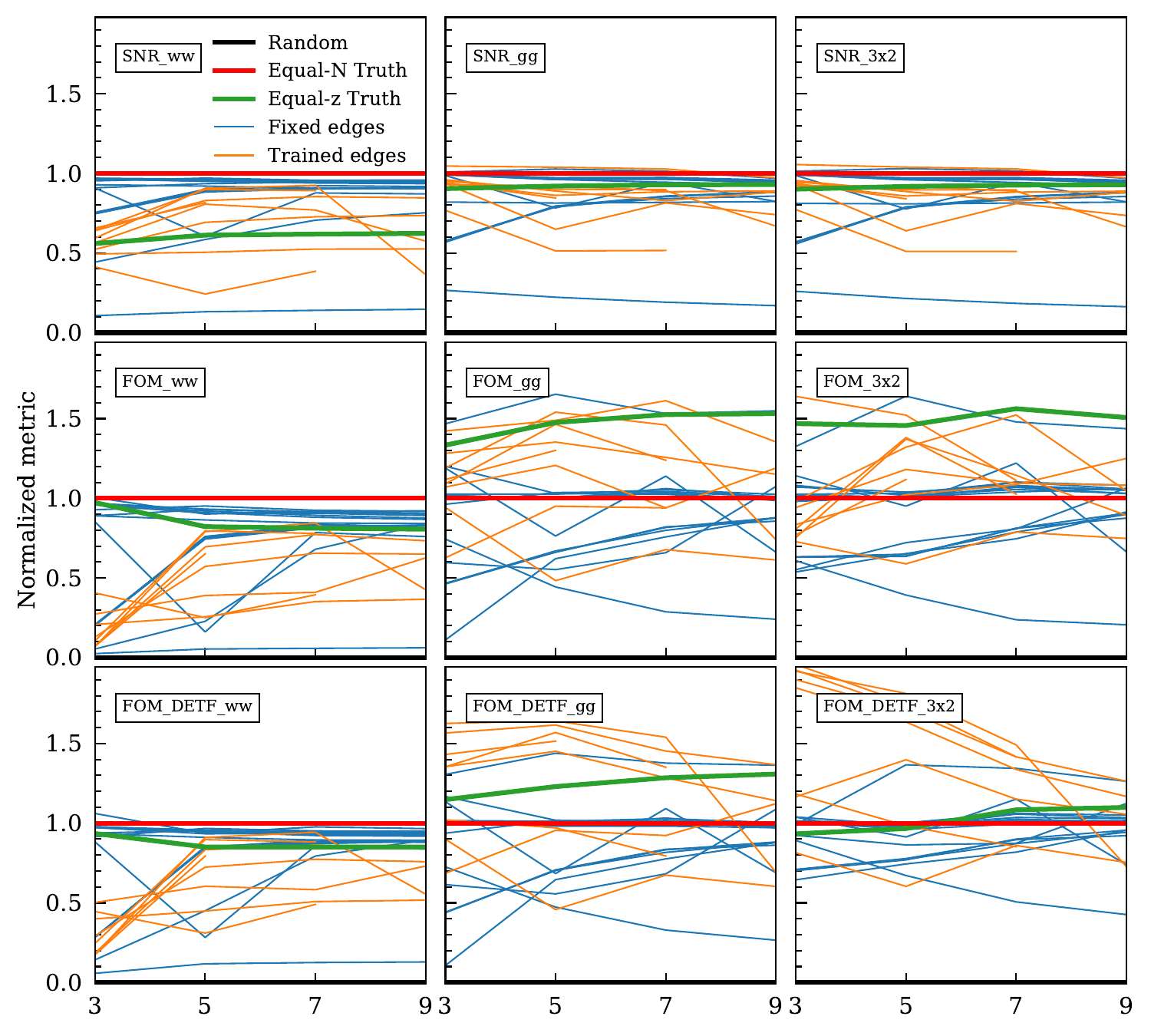}
\caption{Normalized scores for each method, normalized between the score for random assignment (zero) and truth assignment to bins with equal number counts (one).  It can be seen that both equal-number and equal-redshift spacings can be beaten for certain metrics.  In each case, classification is done using griz bands on the CosmoDC2 data set; equivalent plots for Buzzard and riz are shown in \autoref{app:metric_grids}.
\emph{ww} indicates lensing-only metrics, \emph{gg} indicates clustering only, and \emph{3x2} their combination.
The three metrics are defined in \autoref{sec:metrics}.
}
\label{fig:metric_grid_dc2_griz}
\end{figure*}

A comparison of trained-edge and fixed-edge methods on the CosmoDC2 griz scenarios
is shown in Figure \ref{fig:edges}, and the same data normalized against the score achieved
by objects binned in true redshift (i.e. a perfect assigment method) are shown in Figure \ref{fig:metric_grid_dc2_griz}.

One important point to note is that while trained-edge
methods dominate the highest scores for the 3x2pt DETF metric, on which they trained, they
fare much worse on the lensing-only metrics (the methods were not re-trained on the lensing metrics;
we show the metrics for the same assigned bins).  This suggests that the difference between
optimal bins for different science cases is a significant one, and thus that analysis pipelines
should make it straightforward to calibrate multiple configurations and pass them to
subsequent science analysis; this will probably be true even when comparing models
differing only in their systematic parametrizations.  
This feature is highlighted by the comparison between the two variants
of the \textsc{NeuralNetwork} entry, which differed only in which metric they were targeting;
optimizing one metric does not optimize the other (though the changes between results on the two simulations
could also indicate that this is related to the high-redshift behaviour in the CosmoDC2 simulation
described in \ref{sec:cosmodc2}).

Among methods with fixed bins, those using equal numbers in each tend to plateau at scores of
120-140 in Table \ref{tab:cosmodc2}, suggesting that a range of machine learning models can
effectively classify by bin, at least in this case of representative, extensive training
information.  Notably, this is the case even for the \textsc{UTOPIA} entry, which uses the nearest 
training set neighbour to assign bins and thus is explicitly optimised
for this idealised scenario; it should give an upper limit when using the same fixed bins.
That other methods with the same nominal bin choice achieve scores close to it once again
confirms that many entries are close to the best possible score for this approach.

Several of the methods with optimized target bin edges break this
barrier and achieve very high scores.  This included \textsc{ZotNet}, \textsc{ZotBin}, and \textsc{JaxResNet}.  The
\textsc{Stacked Generalization} method similarly included manually tuned bin edges to maximize
score for this scenario (and indeed its ensemble combination method could be applied to
combine the other successful methods). The typical 15--20\% improvement in FoM scores when optimizing
nominal bin edges makes clear that re-selecting edges for each science case is worth the time and effort required.

Notably, though, the winning CosmoDC2 \textsc{FunBins} method used the simpler approach desribed above. The 
success of this simple and fast solution suggests this as an easy heuristic for well-optimized nominal edges, 
at least for cases where clustering-like metrics are important.  Its relatively lower
scores on the Buzzard metrics, as noted in \autoref{sec:metric-results}, does however further highlight
the importance of science- and sample-specific training.

\section{Conclusions} \label{sec:conclusion}
The Tomography Challenge has proved a useful mechanism for encouraging the development
of methods for this science question.  We encourage this challenge structure for other
well-specified science cases, and DESC has subsequently run other challenges along the same lines
such as the N5K beyond-limber challenge\footnote{\url{https://github.com/LSSTDESC/N5K}}.
We describe in Appendix \ref{sec:mistakes} some lessons we have learned from this challenge
that may be useful for future ones.

The results of the challenge have shown that three or four-band photometry
is sufficient for specifying tomography even up to a large number of bins, if
the training data is complete enough: degeneracies between colour do not make this impossible.
Excluding the g-band from the data reduces the constraining power by a noticeable but
not catastrophic amount ($\sim 15$\%).

The wide range in scores for nominally similar methods (for example, the several methods using
neural network approaches) reminds us that, generically, the implementation details of machine
learning methods can drastically affect their performance.

Finally, we have shown that in general there is a signficant advantage to optimizing
target bins afresh when considering a new science case; this can be a significant boost
to the final constraining power.
Despite this, evenly spaced bins in fiducial co-moving distance have proven
a good general choice, as shown in the winning method in the challenge, {\sc FunBins}.

The challenge was heavily simplified, with the intention of testing whether
effective bin-assignments with limited photometry is possible even in theory,
and of generating as many potential methods for the task as possible.  With
both these achieved, future directions for simulating the problem are clear.
The most obvious is limited spectroscopic completeness and size.  Realistic
training data sets will be far smaller than the millions of objects we used here,
and the difficulty of measuring spectra for faint galaxies will make them
incomplete and highly unrepresentative. Machine learning methods will be particularly
affected by the latter problem.  Future challenges must explore these important limitations
using simulated incompleteness, and use realistic photo-z methods to compute metrics.

\section{Acknowledgements}

This paper has undergone internal review in the LSST Dark Energy Science Collaboration.
The internal reviewers were Markus Michael Rau, Christopher Morrison, and Shahab Joudaki.

AHW is supported by a European Research Council Consolidator Grant (No. 770935).
The work of EG and ABr was supported by the U.S. Department of Energy, Office of Science, Office of High Energy Physics Cosmic Frontier Research program under Award Number DE-SC0010008.
CRB, ESC, BMOF, EJG, and GT made use of multi-GPU Sci-Mind Brunelleschi machines developed and tested for Artificial Intelligence and would like to thank CBPF Multi GPU development team of LITCOMP/COTEC, for supporting this infrastructure.
ESC acknowledges support from the FAPESP (\#2019/19687-2) and CNPq  (\#308539/2018-4).
MB acknowledges financial contributions from the agreement ASI/INAF 2018-23-HH.0, Euclid ESA mission - Phase D.
PH acknowledges generous support from the Hintze Family Charitable Foundation through the Oxford Hintze Centre for Astrophysical Surveys.

The DESC acknowledges ongoing support from the Institut National de
Physique Nucl\'eaire et de Physique des Particules in France; the
Science \& Technology Facilities Council in the United Kingdom; and the
Department of Energy, the National Science Foundation, and the LSST
Corporation in the United States.  DESC uses resources of the IN2P3
Computing Center (CC-IN2P3--Lyon/Villeurbanne - France) funded by the
Centre National de la Recherche Scientifique; the National Energy
Research Scientific Computing Center, a DOE Office of Science User
Facility supported by the Office of Science of the U.S.\ Department of
Energy under Contract No.\ DE-AC02-05CH11231; STFC DiRAC HPC Facilities,
funded by UK BIS National E-infrastructure capital grants; and the UK
particle physics grid, supported by the GridPP Collaboration.  This
work was performed in part under DOE Contract DE-AC02-76SF00515

\subsection{Author Contributions}
Zuntz led the project, ran the challenge, wrote most paper text, and made plots.
Lanusse submitted methods, made plots, and contributed to metrics, infrastructure, analysis, and conclusions.
Malz and Wright made plots, and contributed to analysis and conclusions; Wright also submitted methods.
Slosar submitted methods, wrote initial infrastructure, and assisted with challenge design.
Authors Abolfathi to Teixeira submitted methods to the challenge.
Subsequent authors generated the data used in the challenge and contributed to infrastructure used in the project.

Authors CRB, ESC, BMOF, EJG, and GT are collaborators from outside the
Dark Energy Science Collaboration included here under a DESC publication policy exemption.
\vspace{1cm}

\subsection{Author Affiliations}

$^{1}$Institute for Astronomy, University of Edinburgh, Edinburgh EH9 3HJ, United Kingdom\\
$^{2}$AIM, CEA, CNRS, Universit\'e Paris-Saclay, Universit\'e Paris Diderot, Sorbonne Paris Cit\'e, F-91191 Gif-sur-Yvette, France\\
$^{3}$German Centre of Cosmological Lensing, Ruhr-Universit\"{a}t, Universit\"{a}tsstra{\ss}e 150, 44801 Bochum, Germany\\
$^{4}$Brookhaven National Laboratory, Upton NY 11973, USA\\
$^{5}$Department of Physics and Astronomy, University of California, Irvine, CA 92697, USA\\
$^{6}$Astrophysics, University of Oxford, Denys Wilkinson Building, Keble Road, Oxford, OX1 3RH, UK\\
$^{7}$Centro Brasileiro de Pesquisas F\'isicas, Rua Dr. Xavier Sigaud 150, 22290-180 Rio de Janeiro, RJ, Brazil\\
$^{8}$Centro Federal de Educa\c{c}\~{a}o Tecnol\'{o}gica Celso Suckow da Fonseca, Rodovia M\'{a}rcio Covas, lote J2, quadra J - Itagua\'{i} (Brazil)\\
$^{9}$INAF - Osservatorio Astronomico di Capodimonte, via Moiariello 16, I-80131, Napoli Italy\\
$^{10}$Department of Physics and Astronomy, Rutgers, the State University, Piscataway, NJ 08854, USA\\
$^{11}$Universit\'e Paris-Saclay, CNRS/IN2P3, IJCLab, 91405 Orsay, France\\
$^{12}$INFN - Sezione di Napoli, via Cinthia 21, I-80126 Napoli, Italy\\
$^{13}$Department of Physics ``E. Pancini'', University of Naples Federico II, via Cintia, 21, I-80126 Napoli, Italy\\
$^{14}$Universidade de S\~ao Paulo, IAG, Rua do Mat\~ao 1225, S\~ao Paulo, SP, Brazil\\
$^{15}$Instituto de Astronom\'{\i}a Te\'orica y Experimental (IATE-CONICET), Laprida 854, X5000BGR, C\'ordoba, Argentina.\\
$^{16}$Observatorio Astron\'omico de C\'ordoba, Universidad Nacional de C\'ordoba, Laprida 854, X5000BGR, C\'ordoba, Argentina.\\
$^{17}$Dunlap Institute for Astronomy and Astrophysics, University of Toronto, 50 St George St, Toronto, ON M5S 3H4, Canada\\
$^{18}$Department of Astrophysical Sciences, Princeton University, Peyton Hall, Princeton, NJ 08544, USA\\
$^{19}$Department of Physics and Astronomy, University of California, Davis, Davis, CA 95616, USA\\
$^{20}$Argonne National Laboratory, 9700 S Cass Ave, Lemont, IL 60439, USA\\
$^{21}$Department of Physics and Astronomy, Rutgers University, Piscataway, NJ 08854, USA

\bibliography{paper}

\appendix

\section{Theory}\label{app:theory}

In our metrics, the power spectrum between two tomographic bins $i$ and $j$ is
computed using either the Core Cosmology Library \citep{ccl} or the JAX
Cosmology Library \citep{jax-cosmo} as:

\begin{equation}
    C^{ij}_\ell = \int_0^{\infty} \frac{q_i(\chi) q_j(\chi)}{\chi^2} P\left(\frac{\ell +\frac{1}{2}}{\chi}, \chi \right) \mathrm{d}\chi
\end{equation}
where $\chi = \chi(z)$ is the comoving distance and $P$ the non-linear matter power spectrum at
a fiducial cosmology.  The kernel functions $q_i(\chi)$ are different for the lensing and clustering samples:
\begin{align}
    q^{\mathrm{clust}}_i(\chi) &= n_i(\chi)\\
    q^{\mathrm{lens}}_i(\chi) &= \frac{3}{2}\Omega_m \left(\frac{\mathrm{H}_0}{c}\right)^2 \frac{\chi}{a(\chi)} \int_\chi^{\infty} \frac{\chi' - \chi}{\chi'} n_i(\chi')\,\,\mathrm{d}\chi'
\end{align}
where $n_i(\chi) = n_i(z) \frac{\mathrm{d}z}{\mathrm{d}\chi}$.

We evaluate these at 100 $\ell$ values from $\ell_\mathrm{min}=100$ to $\ell_\mathrm{max}=2000$.

We use a Gaussian estimate for the covariance \citep{takada_jain}; this also incorporates the number of galaxies in the
sample:
\begin{equation}
    \mathrm{Cov}(C^{ij}_\ell, C^{mn}_\ell) = \frac{1}{(2 \ell + 1)\Delta\ell f_\mathrm{sky}}(D^{im}_\ell  D^{jn}_\ell + D^{in}_\ell D^{jm}_\ell)
\end{equation}
where $D^{ij}_\ell = C^{ij}_\ell + N^{ij}_\ell$, $\Delta\ell$ is the size in $\ell$ of the binned spectra, and we assume an $f_\mathrm{sky}=0.25$.  The noise spectra are:
\begin{align}
N^{\mathrm{clust},ij}_\ell = \delta_{ij} / n_i \\
N^{\mathrm{lens},ij}_\ell = \delta_{ij} \sigma_{e,\mathrm{tot}}^2 / n_i
\end{align}
where $n_i$ is the angular number density of galaxies in bin $i$ (we asssume equally weighted galaxies, and a 
total number density over all bins of 20 galaxies per square arcminute) and $\sigma_{ei}=0.26$ per shear component, giving $\sigma_{e,\mathrm{tot}} = 0.37$, roughly matching that found in previous surveys \citep{des,desy3a,desy3b,kids,hsc}.

This model does not include a number of theoretical effects that will be important in real data, including
intrinsic alignments, non-Gaussian covariance, and a range of effects on spectra at small scales, like non-linear 
bias and beyond-Limber effects. Collectively these effects mean that our tests are exaggerating the constraining
power of small-scale signals, and increasing all our metrics (especially as our $\ell$ range is fixed with $z$).  
Although this unrealistic aspect mostly affects algorithms equally, the relation between redshift, angle, and physical scale means that the exaggeration is especially strong for low redshift, and so it
could plausibly favour methods that create narrower bins in that range.

\section{Method Details} \label{app:methods}
This appendix details the different methods submitted to the challenge.
Methods described as using optimized bin edges are those that appear as orange solid
lines in \autoref{fig:edges}; those without this note appear as dashed blue lines there.

Method implementations used in the challenge may be found in the \textsc{grand\_merge} branch
of the challenge repository \url{http://github.com/LSSTDESC/tomo_challenge}.

\subsection{ {\sc RandomForest} } \label{sec:randomforest}
This method was submitted by the challenge organizers and used a random
forest algorithm to assign galaxies.

Random forests \citep{breiman2001} train a collection of individual \emph{decision trees},
each of which consists of a set of bifurcating comparisons in which different features in
the data (in this case, magnitudes and errors) are compared to criterion values to choose
which branch to follow.  Each ``leaf'' (end comparison) selects a classification bin for
an input galaxy, and the tree is trained to choose comparison features and comparison values
which maximize discrimination at each step and final leaf purity.

Since decision trees tend to over-fit, random forests generate a suite of 
trees, each randomly choosing a subset of features on which to train at each bifurcation.
An average over the trained trees is taken as the overall classification.

In this implementation we chose nominal bin edges by splitting ensuring equal numbers of training set 
objects were in each bin, and then trained the forest to map magnitudes, magnitude errors, and
galaxy sizes to bin indices.  We use the {\sc scikit-learn} implementation of the random forest
\citep{scikit-learn}.

\subsection{ {\sc LSTM} } 
\label{ClecioLSTM} 
Methods \ref{ClecioLSTM} - \ref{ClecioEnsemble} were
submitted by a team of authors CRB, BF, GT, ESC, and EJG.  All were built using TensorFlow and/or Keras, and
chose bin edges such that an equal number of galaxies is assigned to each bin, and trained the
network using the galaxies' magnitudes and colours. 

The LSTM method uses a Bidirectional Long Short Term Memory network to assign galaxies to redshift
bins.
 
Recurrent Neural Networks (RNNs) \citep{schuster, medsker, pascanu} are a type of NN capable of
analyzing sequential data, where data points have a strong correlation with their neighbours. Long
Short Term Memory (LSTM) units are a particular type of RNN capable of retaining relevant
information from previous points while at the same time removing unnecessary data. This is achieved
through a series of gates with learnable weights connected to different activation functions to
balance the amount of data retained or removed.
 
In some cases, relevant information can come both from data points coming before or after each point
being analyzed. In such cases, two LSTMs can be combined, each going in a different direction, to
create effectively a bidirectional LSTM cell \citep{schuster}.
 
The Deep model is composed of a series of convolutional blocks (1d
convolutional layer with a tanh activation followed by a MaxPooling layer) followed by a
bidirectional LSTM layer and a series of fully connected layers. 

\subsection{ {\sc AutokerasLSTM}}
Autokeras \citep{autokeras} is an AutoML system based on Keras. Given a general neural network
architecture and a dataset, it searches for the best possible detailed configuration for the problem
at hand. 
 
We started from the basic architecture and bin assignment scheme mentioned in Section
\ref{ClecioLSTM} and let Autokeras search for the configuration that results in the lowest
validation loss. We kept the order of the layer blocks fixed, i.e., convolutional blocks,
bidirectional LSTM, and dense blocks. We left the other hyperparameters free, such as the number of
neurons,  dropout rates, maxpooling layers, the number of convolutional filters, and strides.

\subsection{ {\sc LGBM} }
This method uses a Light Gradient Boosted Machine (LightGBM) \citep{lgbm} to assign galaxies to
redshift bins.
 
LightGBM uses a histogram-based algorithm that assigns continuous feature values to discrete bins
instead of other pre-sort-based algorithms for decision tree learning. LightGBM also grows the trees
leaf-wise, which tends to achieve lower losses than level-wise tree growth. This method tends to
overfit for small data so that a \textit{max\_depth} parameter can be specified to limit tree depth.

\subsection{ {\sc TCN}}
This method uses a Temporal-Convolutional network (TCN) \citep{baitcn} to assign galaxies to redshift
bins.
 
Recurrent Neural Networks (such as LSTMs) are considered to be the state-of-the-art model for
sequence modelling. However, research indicates that certain convolutional architectures can achieve
human-level performance in these tasks \citep{dauphin}.
 
TCNs are Fully Convolutional Networks with causal convolutions, in which the output at time
\textit{t} is convolved only with elements of time \textit{t} and lower in the previous layer. This
new approach prevents leakage of information from the future to the past. This simple model has one
disadvantage, in that it needs a very deep architecture or large filters to achieve a long history
size. 
 
By using dilated convolutions, where a fixed step is introduced between adjacent filter taps, the
receptive field of the TCN can be increased without increasing the number of convolutional filters.
Residual blocks \citep{resnet} are also used to stabilize deeper and larger networks. These
modifications cause TCNs to have a longer memory than traditional RNNs with the same capacity,  and
also require low memory for training.
 
We used the Keras implementation of TCN \citep{kerastcn} and chose bin edges such that an equal
number of galaxies is assigned to each bin, and trained the network using the galaxies' magnitudes
and colours. 

\subsection{ {\sc CNN} } \label{sec:cnn}
This method uses an optimized Convolutional Neural Network (CNN) to assign
galaxies to redshift bins. 
 
CNNs \citep{lecun2015deep} are layers inspired in pattern recognition types developed in mammals' brains.
They consist of kernels that are convolved with the data as it flows through the net. CNN-based
models have emerged as state-of-the-art in several computer vision tasks, such as image
classification and pose estimations, as well as having applicability in a range of different fields.
 
We combine the convolutional layers with dimensionality reduction (pooling layers) and different
activation functions. Here, we used Autokeras to optimize a general convolutional architecture,
using the galaxies' magnitudes and colours to assign them to redshift bins, which were chosen such
that an equal number of galaxies were assigned to each.

\subsection{ {\sc Ensemble} } 
\label{ClecioEnsemble} 
This method combines different neural network
architectures to assign galaxies to redshift bins.
 
Deep Ensemble models combine different neural network architectures to obtain better performance
than any of the nets alone. In this method, we used the Bidirectional LSTM optimized by Autokeras, a
ResNet \citep{resnet} and a Fully Convolutional Network (FCN) \citep{fcn}. The predictions of these
models were averaged to get the final predicted bin.

\subsection{ {\sc SimpleSOM} }
The \textsc{SimpleSOM} algorithm submitted by author AHW utilises a self-organising map
\citep[SOM,][]{Kohonen:1982} to perform  a discretisation of the
high-dimensional colour-magnitude-space of the challenge training and reference
datasets. 

We utilise a branched version of the {\sc
kohonen}\footnote{\url{https://github.com/ AngusWright/kohonen.git}} package
\citep{Wright/etal:2020b, Wehrens/Kruisselbrink:2018, Wehrens/Lutgarde:2007}
within R \citep{R}, which we integrate with  the challenge's python
classes using the {\sc rpy2} interface\footnote{\url{https://rpy2.github.io}}. 
We train a $101 \times 101$ hexagonal-cell SOM with toroidal topology on the
maximal combination of (non-repeated) colours, $z$-band magnitude, and so-called
`band-triplets'. For example, for the `griz' setup, our SOM is trained on the
combination of  $z$, $g-r$, $g-i$, $g-z$, $r-i$, $r-z$, $i-z$, $g-r-(r-i)$,
$r-i-(i-z)$. This combination proved optimal during testing on the DC2 dataset,
as determined by the 3x2pt SNR metric. The training and reference data then
propagated into this trained SOM, producing like-for-like groupings between the
two datasets. We then compute the mean redshift of the training sample sources
within each cell and the number of reference sources per-cell. By rank-ordering
the cells in mean training-sample redshift, we construct the cumulative
distribution function of reference sample source counts, and split cells into
${n_{\rm tomo}}$ equal-N tomographic bins using this function. 

In addition to this base functionality, the \textsc{SimpleSOM} method includes a
mechanism for distilling $N_{\rm cell}$ SOM cells into $N_{\rm group}<N_{\rm
cell}$ groups of cells, while maintaining optimal redshift resolution. This is
done by invoking a hierarchical clustering of SOM cells \citep[see Appendix B of
][]{Wright/etal:2020a}, where in this case we combine cells based on the
similarity of the cells' $n(z)$ moments (mean, median, and normalized median absolute
deviation from median) using complete-linkage clustering. The result of this
procedure is the construction of fewer discrete groups of training and reference
data, while not introducing pathological redshift-distribution broadening (as
happens when, e.g., training using a lower resolution SOM and/or clustering
cells based on distances in colour-magnitude space). 

\subsection{PQNLD} \label{sec:pqnld}

The `por que no los dos' algorithm, submitted by author AHW, is a development of the \textsc{SimpleSOM} algorithm, adding the 
additional complexity of also computing template-based photometric redshift point-estimates using BPZ 
\citep{Benitez:2000}. Photo-$z$ point-estimates are derived using the re-calibrated 
template set of \cite{Capak:2004} in combination with 
the Bayesian redshift prior from \cite{Raichoor/etal:2014}, using only the (g)riz bands supplied in the challenge. 

The algorithm seeks to leverage information from both template-based 
photometric redshift estimates and machine-learning to improve the allocation of discretised cells in 
colour-magnitude-space to tomographic bins, by flagging and discarding sources which reside in cells that have 
significant colour-redshift degeneracies. In practice this is achieved by performing a quality-control step prior to 
the assignment of SOM cells to tomographic bins, whereby each cell $i$ is flagged as having a catastrophic 
discrepancy between its mean training $z$ and mean photo-$z$ if: 
\begin{equation}
\frac{|\langle z_{\rm train,i}\rangle - \langle z_{\rm phot,i}\rangle  -\langle z_{\rm train} - z_{\rm phot}\rangle|}
{\sigma[z_{\rm train} - z_{\rm phot}]} \geq 5.
\end{equation}

This flagging is inspired by similar quality control that is implemented in the redshift calibration procedure of
\citet{Wright/etal:2020a}, which removes pathological cells in the SOM prior to the construction of calibrated 
redshift distributions for KiDS. We note, though, that this procedure is unlikely to have a significant influence in this 
challenge, as it is primarily designed to identify and remove cells with considerable covariate shift between the 
training and reference samples; a problem which does not exist (by construction) in this challenge.

For a further discussion of the merits of band triplets in machine learning photo-z see \citet{broussard_gawiser}

\subsection{ {\sc ComplexSOM} } \label{sec:csom}
This method was submitted by authors DA, AHW, AN, EG, BA, and ABr. 
It used optimized bin edges.

The {\sc ComplexSOM} method implements an additional optimization layer on the
methodology used by {\sc SimpleSOM}.

Let ${\sf C}^{\rm group}_\ell$ be the matrix containing all auto- and
cross-power spectra between members of a large set of galaxy samples (called
``groups'' here). We can compress this large set into a smaller set of samples
(called ``bins'' here), by combining different groups. Let ${\sf A}$ be the
assignment matrix determining these bins, such that $A_{\alpha,i}=1$ if group
$i$ is assigned to bin $\alpha$, and 0 otherwise. In that case, the redshift
distribution of bin $\alpha$ is given by $N_\alpha(z)=\sum_i
A_{\alpha,i}n_i(z)$, where $n_i(z)$ is the redshift distribution of group $i$.

Defining the normalized assignment matrix ${\sf B}$ as
\begin{equation}
B_{\alpha,i}=A_{\alpha,i}\frac{\int dz\,n_i(z)}{\sum_j A_{\alpha,j}\int dz\,n_j(z)},
\end{equation}

${\sf C}^{\rm group}_\ell$ is related to the matrix of cross-power spectra
between pairs of bins via:
\begin{equation}
{\sf C}^{\rm bin}_\ell={\sf B}\,{\sf C}^{\rm group}_\ell\,{\sf B}^T.
\end{equation}

The rationale behind the {\sc ComplexSOM} method is based on the observation
that, while ${\sf C}^{\rm group}_\ell$ is a slow quantity to calculate for a
large number of groups, ${\sf B}$ and therefore ${\sf C}^{\rm bin}_\ell$ are
very fast. Thus, given an initial set of groups characterizing the
distribution of the full sample in colour space, ${\sf C}^{\rm group}_\ell$
can be precomputed once, and used to find the optimal assignment matrix ${\sf
B}$ that maximizes the figure of merit in Equation \ref{eq:fom},
which can be written in terms of ${\sf C}^{\rm bin}_\ell$ alone as

\begin{equation}\label{eq:fom_tr}
F_{\theta\theta'}=\sum_\ell f_{\rm sky}\frac{2\ell+1}{2}{\rm Tr}\left[\partial_\theta{\sf C}^{\rm bin}_\ell({\sf C}^{\rm bin}_\ell)^{-1}\partial_{\theta'}{\sf C}^{\rm bin}_\ell({\sf C}_\ell^{\rm bin})^{-1}\right].
\end{equation}

Note that the general problem of finding the matrix ${\sf B}$ that optimally
compresses the information on a given parameter has an analytical solution in
terms of Karhunen-L\`oeve eigenvectors \citep{astro-ph/9603021}. However, the
assignment matrix used in this case has the additional constraints that it
must be positive, discrete and binary (i.e. groups are either in a bin or they
are not). Thus, finding the absolute maximum figure of merit would involve a
brute-force evaluation of all matrix elements $A_{\alpha,j}$, which is an
intractable $N_{\rm bin}^{N_{\rm group}}$ problem. We thus parametrize ${\sf
A}$ in terms of a small set of continuous parameters, for which standard
minimization routines can be used. In particular we use $N_{\rm bin}-1$
parameters, given by the edges of the redshift bins, and a group is assigned
to a bin if its mean redshift falls within its edges. We also explored the
possibility of ``trashing'' groups if less than a fraction $p_{\rm in}$ of its
redshift distribution fell within its preassigned bin, treating $p_{\rm in}$
as an additional free parameter. We found that adding this extra freedom did
not translate into a significant improvement in the final level of data
compression.

Once the SOM has been generated, the {\sc ComplexSOM} method proceeds as
follows. First, the full SOM is distilled into a smaller set of ${\cal O}(100)$
groups based by combining SOM cells with similar moments of their redshift
distributions, thus ensuring that this process does not induce any catastrophic
$N(z)$ broadening.
The large ${\sf C}^{\rm group}_\ell$ is then precomputed for the initial set of
groups. Finally, we find the optimal assignment matrix, defined in terms of the
free redshift-edge parameters, by maximizing the figure of merit in Eq.
\ref{eq:fom_tr} using Powell's minimization method
\citep{10.1093/comjnl/7.2.155}.

\subsection{ {\sc UTOPIA} }\label{sec:utopia}

The `Unreliable Technique that OutPerforms Intelligent Alternatives' is a method
submitted by AHW that is designed,  primarily, as a demonstration of the care
that must be taken in the interpretation of the tomographic challenge results. 
The method performs  the simplest-possible direct mapping between the training
and reference data, by performing a nearest-neighbour  assignment of training
sample galaxies to each reference sample source in $n_{\rm band}$ magnitude-only
space. Each reference  sample galaxy is then assigned the redshift of its
matching training-sample source, and these redshifts are  used to bin the
reference galaxies into $n_{\rm tomo}$ equal-N tomographic bins. Importantly,
the method was specifically  implemented so as to use only the minimum
information contained within the available bands (by using magnitudes alone,
without  any colours, etc.). 

In this way \textsc{UTOPIA} is the simplest possible algorithm (utilising all available
photometric bands) that one can  implement for tomography. It is also a method
whose performance is optimal in the limit of an infinitely large,  perfectly
representative training sample; i.e. the limit in which this challenge operates.
As the requirements of perfect  training-sample representivity and large size
are violated, \textsc{UTOPIA} ought to assign redshifts to reference  galaxies from
increasingly distant regions of the multi-dimensional magnitude-space, resulting
in poorer (i.e. unreliable) tomographic binning.  Therefore, \textsc{UTOPIA} acts as a warning against
over-interpretation of the tomographic challenge results, as success in the
challenge  need not translate to optimal-performance/usefulness under realistic
survey conditions. Instead, the challenge results should be interpreted
holistically, by comparing the performance between different classes of methods,
rather than individuals. The authors have endeavoured to do so in the conclusions presented here.

\subsection{ {\sc GPzBinning} }
This method was submitted by author PH.

GPzBin is based on GPz \citep{Almosallam2016a,Almosallam2016b}, a machine learning
regression algorithm originally developed for calculating the photometric
redshifts of galaxies \citep{Gomes2017,Duncan2018,Hatfield2020} but now also
applied to other problems in physics \citep{Peng2019,Hatfield2020}. The
algorithm is based on a Gaussian Process (GP; essentially an un-parametrised continuous
function defined everywhere with Gaussian uncertainties); GPz specifically uses 
a sparse GP framework, a fast and a scalable approximation
of a full process \citep{Rasmussen2006}.

The mean $\mu(x)$ and variance $\sigma(x)^2$ for predictions as a function of
the input parameters $x$ (typically the photometry) are both linear combinations
of a finite number of `basis functions': multi-variate Gaussian kernels with
associated weights, locations and covariances. The algorithm seeks to find the
optimal parameters of the basis functions such that the mean and variance
functions are the most likely functions to have generated the data. The key
innovations introduced by GPz include a) input-dependent variance estimations
(heteroscedastic noise), b) a `cost-sensitive learning' framework where the
algorithm can be tailored for the precise science goal, and c) properly
accounting for uncertainty contributions from both variance in the data
($\beta^{-1}_{\star}$) as well as uncertainty from lack of data ($\nu$) in a
given part of parameter space (by marginalising over the functions supported by
the GP that could have produced the data).

GPzBin is a simple extension to GPz for the problem of tomographic binning. For
a given number of tomographic bins it selects redshift bin edges $z_{i}$ such
that there are an equal number of training galaxies in each bin. Testing sample
galaxies are then assigned to the bin corresponding to their GPz predicted
photometric redshift. The code allows for two selection criteria for removing
galaxies for which it was not possible to make a good photometric redshift
prediction: 1) cutting galaxies close to the boundaries of bins based on an
`edge strictness' parameter $U$ which removes galaxies where $\mu\pm U \times
\sigma \lessgtr z_i$ and 2) cutting galaxies based on the degree $E$ to which
GPz is having to extrapolate to make the redshift prediction, removing galaxies
with $\nu>E\sigma^2$ (see Figure 12 of \citealp{Hatfield2020}). Cutting galaxies
removes some signal at the possible benefit of improving binning purity. As the
training and test data were sampled from the same distribution in this data
challenge we might not expect removing galaxies to give huge improvements in
binning quality, but cuts based on $E$ and $U$ might become more useful in
future studies where the training and test data have substantially different
colour-magnitude distributions.

We used these settings in GPz for this challenge (see
\citealp{Almosallam2016a,Almosallam2016b} for definitions): x=100, maxIter=500, 
maxAttempts$=50$, method=GPVC, normalize=True, and joint=True.

\subsection{{\sc JaxCNN \& JaxResNet} }
 
These methods were submitted by the author AP.
They used optimized bin edges.
 
\textsc{JaxCNN} and \textsc{JaxResNet} are convolutional neural network based algorithms that
map the relationship between the colour-magnitude data of galaxies to their bin
indices. Convolutional neural network (CNN) models, as described in section
\ref{sec:cnn}, consist of several layers, each consisting of a linear
convolution operator followed by polling layers and a nonlinear activation
function. Residual Network (ResNet) models take one step further and learn the
\textit{identity} mapping by skipping, or adding \textit{shortcut} connections
to, one or more layers \citep{resnet}. ResNet helps solve the vanishing gradient
problem, allowing a deeper neural network.
 
\textsc{JaxCNN} uses 2-layer convolutional neural networks and \textsc{JaxResNet} uses a stack of
3 layers consisting of $1 \times 1, 3 \times 3$, and $1 \times 1$ convolutions
with shortcut connections added to each stack. Both methods use the Rectified
Linear Unit (ReLU) activation function under the ADAM optimizer \cite{adam} with
a learning rate of 0.001. 
 
We chose the learning algorithm for both methods to minimize the reciprocal of
the figure of merit $F$, where $F$ is defined in \autoref{eq:fom}. We implement
these models using the JAX-based Flax neural network library \cite{jax}.

\subsection{ {\sc NeuralNetwork 1 \& 2} } \label{sec:nn}
These methods were submitted by the author FL.
They used optimized bin edges

The {\sc NeuralNetwork} approach combines the two following ideas:
\begin{itemize}
	\item Using a simple neural network to parameterize a bin assignment function $f_\theta$ taking available photometry as the input and returning a bin number. 
	\item Optimizing the parameters of $f_\theta$ as to maximize a target score, either the total SNR (\autoref{eq:snr}), or the DETF FoM (\autoref{eq:fom}).
\end{itemize}
The method directly solves the problem of optimal bin assignment given available photometry, to maximize the cosmological information, without estimating a redshift.

Details of the particular neural network are mostly irrelevant; the main difficulty in this approach is to be able to compute the back-propagation of gradients through the cosmological loss function to update the parameters $\theta$ of the bin assignment function. This is was made possible for this challenge by the \textsc{jax-cosmo} library \citep{jax-cosmo}, which allows the computation of both metrics using the JAX automatic differentiation framework \citep{jax}.

In practice, to parameterize $f_\theta$ we use a simple dense neural network, composed of 3 linear layers of size 500 with leaky\_relu activation function and output batch normalization. The last layer of the model is a linear layer with an output size matching the number of bins, and softmax activation. We implement this model using the JAX-based \textsc{Flax} neural network library \citep{flax2020github}. 

Training is performed using batches of 5000 galaxies, over 2000 iterations under the ADAM optimizer \citep{adam} with a base learning rate of 0.001. The loss functions $\mathcal{L}$ used for training directly derive from the challenge metrics, and constitute the only difference between the \textsc{NeuralNetwork1} \& \textsc{neurNetwork2} entries:

\begin{itemize}
	\item \textsc{NeuralNetwork1}: $\mathcal{L}_1 = 1 / S_{FoM}$
	\item \textsc{NeuralNetwork2}: $\mathcal{L}_2 = - S_{SNR} $
\end{itemize}

\subsection{ {\sc PCACluster} }
This entry was submitted by author DG.
It used optimized bin edges.

Principal Component Analysis (PCA) reduces the dimensionality of a dataset by
calculating eigenvectors that align with the principal axes of variation in the
data \citep{doi:10.1098}. {\sc PCACluster} uses PCA to reduce the flux \& colour
data set to three dimensions. 

To generate the PCA dimensions used by this method the PCA algorithm takes in
the $r$-band flux and $ri$ and $iz$ colours and outputs three principal
components.  If the data set provides the $g$-band, the algorithm also uses the
$gr$ colour when determining eigenvectors. Tomographic bins are then determined
in this three dimensional space using a clustering algorithm where each
observation is assigned to a bin by determining which cluster centroid is
closest to that observation.

In order to determine optimal centroid positions, clustering is framed as a
gradient descent problem that seeks to maximize the requested metric: either the
SNR or the FoM. This approach of identifying clusters using gradient descent is
inspired by the similar approach to K-means clustering in
\citet{NIPS1994_a1140a3d}.

This entry uses {\sc jax\_cosmo} to take the derivative of the requested metric
and classification function combination with respect to the centroid locations
to calculate gradients and weight updates \citep{jax-cosmo}.  The speed of
training PCACluster is directly proportional to the number of requested
centroids (the number of requested tomographic bins) and the amount of training
data used.

\subsection{ {\sc ZotBin and ZotNet} } \label{sec:zot}

This pair of related methods was submitted by author DPK and its associated code is available at
\url{https://github.com/dkirkby/zotbin}.
It used optimized bin edges.

The ZotBin method consists of three stages: preprocessing, grouping, and bin optimization.  The ZotNet method
uses the same preprocessing, then skips directly to bin optimization.  Both methods optimize the final bins to
maximize an arbitrary linear combination of the three metrics defined in \ref{sec:metrics}, using the JAX
library~\cite{jax} for efficient computation. The goal of ZotNet is to find a nearly optimal set of bins (for
a specified linear combination of metrics) using a neural network, at the expense of interpretability of the
results.  The ZotBin method aims to perform almost as well as ZotNet, but with a much more interpretable mapping
from input features to final bins.

Both methods transform the $n$ input fluxes into $n-1$ colours and one flux. The data are very sparse in this
$n$-dimensional feature space and exhibit complex correlations. The preprocessing step transforms to a new
$n$-dimensional space where the data is nearly uncorrelated and dense on the unit hypercube $[0,1]^n$. This is accomplished
by learning a normalizing flow~\cite{2019arXiv191202762P} that maps the complex input probability density
into a multivariate unit Gaussian, then applying the error function to yield a uniform distribution. Although
the resulting transformation is non-linear, it is invertible and differentiable, and thus provides an interpretable
smooth mapping.

The second stage of ZotBin starts by dividing the unit hypercube $[0,1]^n$ from the preprocessing step
into a regular lattice of ${\cal O}$(10K) cells which, by construction, each contain a similar number
of training samples.  Next, adjacent cells are iteratively merged into $M \sim 100$ groups of cells by
combining at each step the pair of groups that are most ``similar''.  We define similarity as the product
of similarities in feature and redshift space.  Redshift similarity is based on the histogram of redshifts
associated with each group, interpreted as a vector.

The final stage of ZotBin selects linear combinations of the $M$ groups to form the $N$ output bins. This is
accomplished by optimizing the specified metric combination with respect to the $M\times N$ matrix of linear
weights.

The ZotBin method defines a fully invertible sequence of steps that transform the input feature space into
the final bins, so that each bin is associated with a well-defined region in colour / flux space. The ZotNet method,
on the other hand, trains a neural network to directly map from the preprocessed unit hypercube $[0,1]^n$ into
$N$ output bins by optimizing the specified metric combination. The resulting map is not invertible, so less
interpretable, but also less constrained than ZotBin, so expected to achieve somewhat better performance.

\subsection{ {\sc FunBins} } \label{sec:funbins}
This entry was submitted by author ABa.

This method uses the random forest algorithm described in section \ref{sec:randomforest} to assign galaxies to tomographic bins, but modifies the target bin edge selection method. There are 3 options for determining the edges used to assign labels to the training data: \textit{log}, \textit{random}, and \textit{chi}. 

The first option, \textit{log}, calculates bin edges such that the number of galaxies in each bin grows logarithmically instead of being constant as in section \ref{sec:randomforest}. This option uses increasingly large bins for more distant galaxies and tests the relative importance of nearby galaxies.

The second option, \textit{random}, draws the intermediate bin edges from a uniform distribution while fixing the outer edges to the limits of the data. This option is not theoretically motivated and designed only to study the sensitivity to the binning. 

The last option, \textit{chi}, calculates redshift bin edges whose corresponding comoving distances are equally spaced, using Planck15 \citep{Planck15} cosmology for the distance-redshift relationship. This should lead to roughly equal shot noise for the angular power in each bin. 
This method was used in the scores shown in this paper.

Once the bin edges are fixed using either of the methods above, the random forest is trained as described in section \ref{sec:randomforest}.

\subsection{FFNN}
This entry was submitted by author EN.

The method is a Feed Forward Neural Network (FFNN), a classic type of neural network.
This implementation used the Keras API in the TensorFlow library. Three components, a flattener, and two 
dense layers of ReLu neurons, were optimized by ADAM on the sparse categorical cross-entropy of the classifications.
A StandardScaler was used to pre-process the data, and the training stopped when there was no improvement in the  
validation loss for three consecutive epochs, to avoid overfitting.

The network is trained using the galaxies' magnitudes, colours and band-triplets as well as their errors to assign them to 
the redshift bins.
Prior to processing the training and validation data, the non-detection placeholder magnitudes (30.0) were replaced with 
an approximation of the $1\sigma$ detection limit as an estimate for the sky noise threshold, where Signal to Noise ratio, 
S/N, equals 1. We first compute the error equivalent of $dm = 2.5 \log(1 + N/S)$ where $dm \sim 0.7526$ 
magnitudes for $N/S = 1$, and then fit a logarithmic model to the magnitude as a function of its error 
for different bands in the training data. This gives us the magnitude corresponding to this limit 
and in order to be consistent we use this value along with its error to replace the non-detections 
everywhere during the process.

\subsection{ {\sc MLPQNA} }
This entry was submitted by authors SC and MB.

The Multi Layer Perceptron trained by Quasi Newton Algorithm 
({\sc MLPQNA}; \citealp{Brescia12}), is a feed-forward neural network 
for multi-class classification and single/multi regression use cases.

The architecture is a classical Multi Layer Perceptron (MLP; \citealp{Rosenblatt1961})
with one or more hidden layers, on which the supervised learning paradigm is run
using the Quasi Newton Algorithm (QNA), implemented through the L-BFGS rule
\citep{Nocedal80}. L-BFGS is a solver that approximates the Hessian matrix,
representing the second-order partial derivative of a function. Further, it
approximates the inverse of the Hessian matrix to perform parameter updates.
MLPQNA uses the least square loss function with the Tikhonov regularization
\citep{Tikhonov77} for regression. It uses the cross-entropy \citep{deBoer05}
and softmax \citep{Sutton98} as output error evaluation criteria for
classification.

Besides several scientific contexts and projects in which MLPQNA has succesfully
been used, it was recently used as the kernel embedded into the method {\sc
METAPHOR} \citep{cavuoti20} for photo-z estimation, which participated in the
LSST Photometric Redshift PDF Challenge \citep{schmidt20}. 

The {\sc MLPQNA} python implementation used here
uses a customized Scipy version of the L-BFGS built-in rule.
It used two hidden layers with $2N-1$ and $N+1$ nodes respectively,
where $N$ is the number of input features (depending on the experiment).
The decay was set to 0.1.

\subsection{ {\sc Stacked Generalization} } \label{sec:stackgen}
This method was submitted by author JEC.

\emph{Stacked generalization} consists of stacking the output of several individual estimators and 
using a classifier to compute the final prediction. Stacking methods use the strength of each 
individual estimator by using their outputs as input of a final estimator.  This entry stacked
\emph{Gradient Boosted Trees} (GBT) classifiers with standard Random Forests (RFs), and then used a Logistic 
Regression as the final step.

A GBT is a machine learning method which combines decision trees (like the standard RF)
by performing a weighted mean average of individual trees \citep{Friedman:2002we,RefWorks:1634}. 
The prediction $F(X)$ of the ensemble of trees for a new sample $X$ is given by
\begin{equation}
F(X) = \sum_k w_k F_k(X)
\end{equation}
where $F_k(X)$ is the individual tree answer and $w_k$ a weight per tree. Unlike the RF, the 
tree depth is rather small and the training  of the $k$-th tree is based on the errors of the
$(k-1)$-th tree.  Like the RF, a randomly chosen fraction of features ($\approx 50\%$) is 
omitted at each split, and this step is changed at each tree (leading to a Stochastic Gradient Descent 
algorithm).  GTB learning is essentially non-parallelizable and so cannot use a map-reduce paradigm, 
unlike RF. It suffers from the opposite problem to RF's overfitting, and is subject
to high bias, so one generally combines many weak individual learners.

The two methods have their own systematics; they notably differ in the feature importances. Stacking 
the two is therefore especially effective.  The submitted method combined 50 GBTs with
50 RFs using the {\sc SKLearn} {\sc StackingClassifier}.

An iterative precedure was used to progressively change the target bin edges to optimize one of the metrics (FOMs or SNRs) between the trainings of the classifiers. In general, we recover different choices for each statistic, but the differences are relatively marginal. This procedure was done by hand for this challenge, but could be automated and applied to any kind of classifier.

Although this entry stacked GBTs and RFs, the same approach can be used more generally, such as using
one classifier optimized for high redshift and another for low, or one classifier for an SNR metric
and a second for an FOM metric.  This makes stacking approaches particularly flexible.

\section{Mistakes} \label{sec:mistakes}
The challenge organizers made a number of mistakes when building and running the challenge.
These do not invalidate our process or results, but we describe some of them here in the hope of
being useful for future challenge designers.

\subsection{Splitting nominal bin choice from assignment}
As noted in the main test, there were two ways to obtain good scores in the challenge: 
either a team could choose the best
nominal bin edges, or find the best method to put galaxies into those bins once they are chosen.
These two are not completely disconnected: some theoretically powerful choices of nominal edges
could prove impossible to assign in practice.  But it would still have been useful to separate
the challenge into two separate parts, fixing the edges while optimizing classifiers, and vice
versa.   We can retrospectively determine this since we understand the classifiers, but the picture
would have been clearer doing so in advance.

\subsection{Infrastructure}
Since submissions could use a wide variety of libraries and dependencies, it was extremely
difficult after the challenge was complete to build appropriate environments for all the challenges
in order to run the methods on additional data sets. In retrospect, a robust and careful
continuous integration tool with a mechanism for entrants to specify an environment (for example, a 
container or a {\sc conda} requirements file) could have been used to simplify this and ensure
all methods could be easily run by the organizers.

One additional issue, though, was that many methods required GPUs to train in a reasonable time,
and specifying an environment on these systems can be more complicated.

\subsection{Guidance on caching \& batch processing}
Since the training process for the challenge could be slow for some algorithms, many entries
(very reasonably) cached trained models or other values for later use.  In some cases these
cached values were then incorrectly picked up by subsequent runs on new data sets, leading
to (at best) crashes and at worst silently poor scores\footnote{We checked for this problem in 
our final results by running each method and checking for new files in the working
directory.}.  
We should have provided an explicit
caching mechanism for entrants to avoid such issues.

Additionally, the challenge supplied the data described in Section \ref{sec:data} to entrants,
and some assumed that it was safe to hard-code specific paths to them or
otherwise assume fixed data inputs.  This led to problems when switching to new test data sets. 
Again, a requirement enforced by continuous integration could have checked this.

\section{Normalized metric scores} \label{app:metric_grids}
\begin{figure*}
\includegraphics[width=1\linewidth]{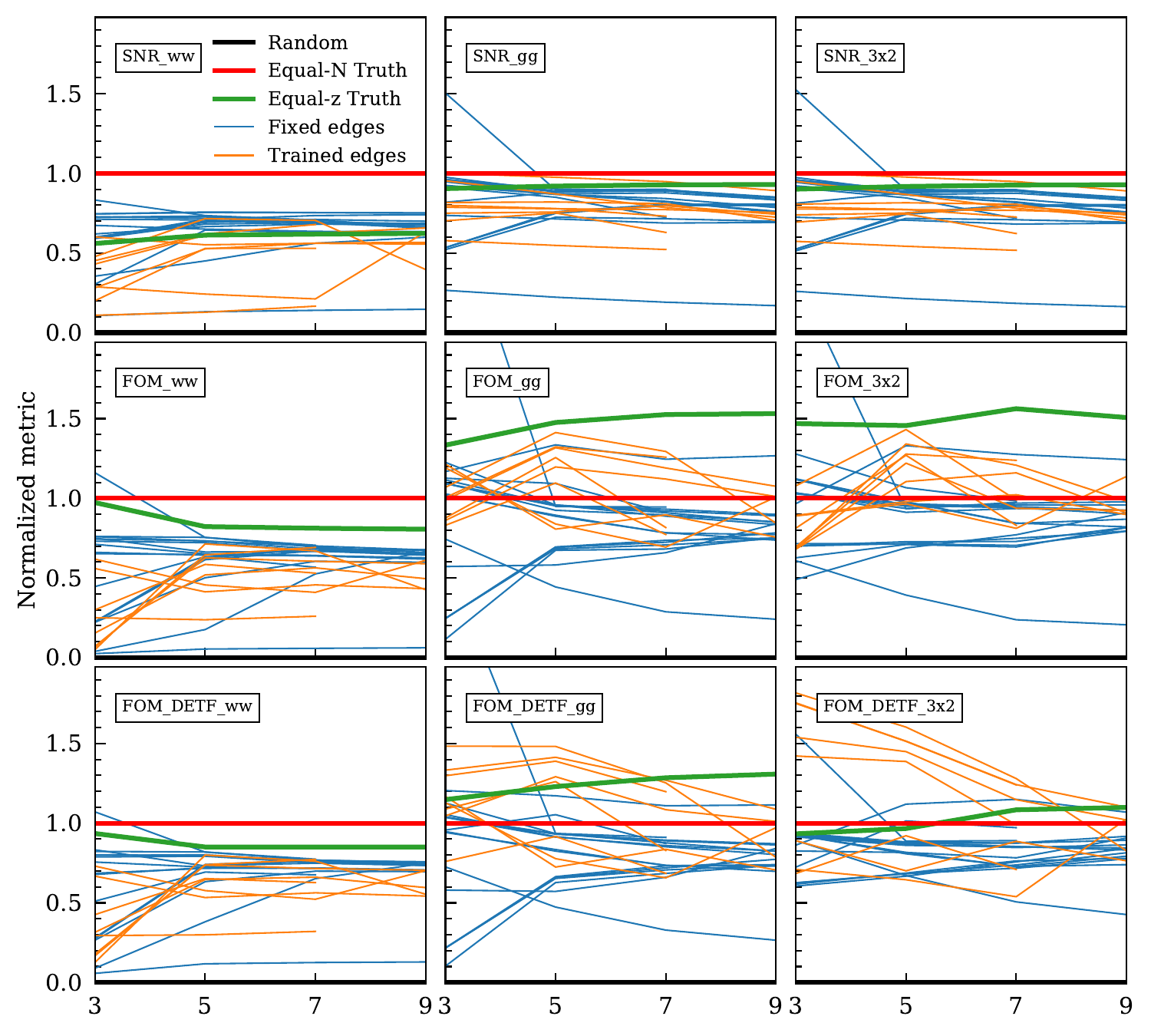}
\caption{Normalized scores for each method, following the scheme of Figure \ref{fig:metric_grid_dc2_griz}
but for CosmoDC2 with riz bands}
\label{fig:metric_grid_dc2_riz}
\end{figure*}

\begin{figure*}
\includegraphics[width=1\linewidth]{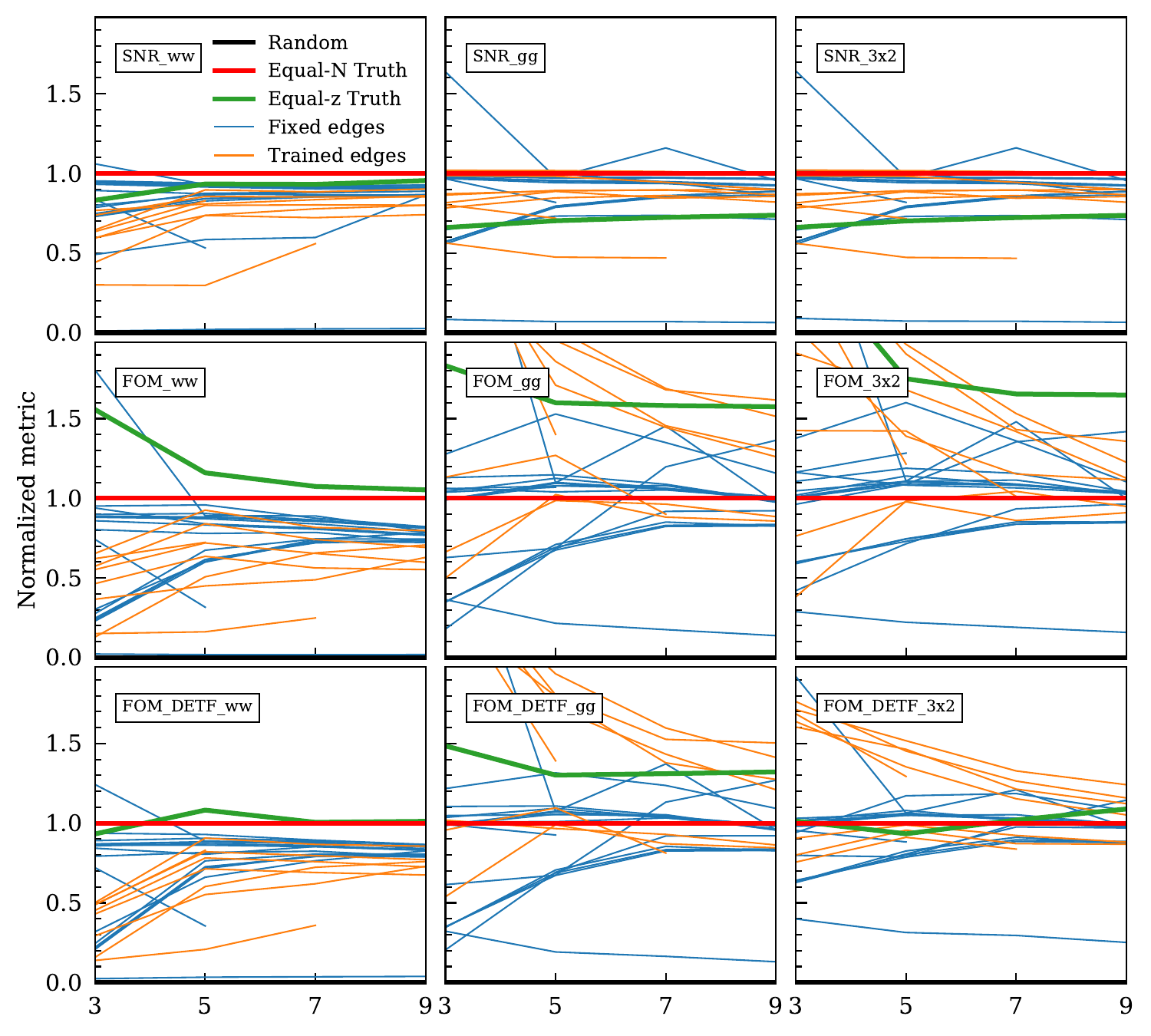}
\caption{Normalized scores for each method, following the scheme of Figure \ref{fig:metric_grid_dc2_griz}
but for Buzzard with griz bands}
\label{fig:metric_grid_buzzard_griz}
\end{figure*}

\begin{figure*}
\includegraphics[width=1\linewidth]{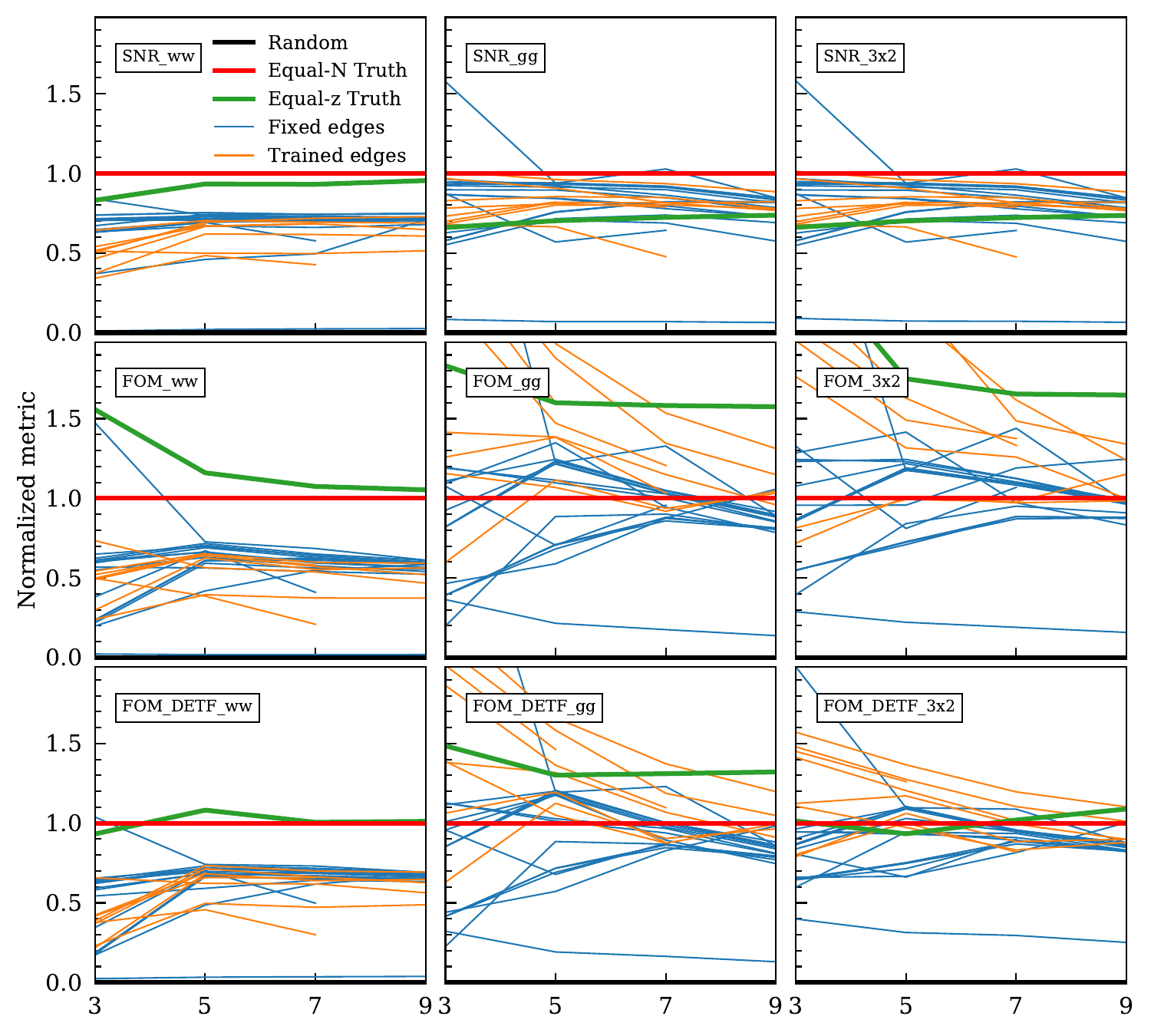}
\caption{Normalized scores for each method, following the scheme of Figure \ref{fig:metric_grid_dc2_griz}
but for Buzzard with riz bands}
\label{fig:metric_grid_buzzard_riz}
\end{figure*}

Figures \ref{fig:metric_grid_dc2_riz}, \ref{fig:metric_grid_buzzard_griz}, and \ref{fig:metric_grid_buzzard_riz} show metric scores normalized between random assignment and perfect assigment to equal number density bins for CosmoDC2/riz, Buzzard/griz and Buzzard/riz respectively:
\begin{equation}
S_\mathrm{norm} = \frac{S - S_\mathrm{random}} {S_\mathrm{perfect,eq} - S_\mathrm{random}}
\end{equation}

\section{Full 9-bin results tables} \label{app:tables}
Tables \ref{tab:full_dc2} and \ref{tab:full_buzz} show all calculated metrics when generating nine tomographic 
bins for CosmoDC2 and Buzzard respectively. The \textit{ww} columns
show weak-lensing metrics, \textit{gg} show galaxy clustering, and \textit{3x} the full 3x2pt metric.
Methods above the horizontal line trained bin edges; methods below used fixed fiducial ones, though in some
cases did some hand-tuning of them before submission. In each section the highest scoring method is highlighted.

As in table \ref{tab:cosmodc2}, some values are missing due to failure or time-out of the method (*) or 
pathological bin assignments (-).  Failure rates were higher for the 9-bin runs than for the lower bin counts.
\vspace{0.5cm}

\renewcommand{\arraystretch}{1.8}
\begin{sidewaystable}
\vspace{10cm}
\begin{tabular}{ |l | c c c | c c c | c c c | c c c | c c c | c c c |}
\hline
\multirow{3}{*}{Method} & \multicolumn{9}{|c|}{riz} & \multicolumn{9}{c|}{griz} \\
\cline{2-19}
&  \multicolumn{3}{c|}{SNR}   &  \multicolumn{3}{c|}{$\sigma_8-\Omega_m$ FOM} & \multicolumn{3}{c|}{$w_0-w_a$ FOM}
&  \multicolumn{3}{c|}{SNR}   &  \multicolumn{3}{c|}{$\sigma_8-\Omega_m$ FOM} & \multicolumn{3}{c|}{$w_0-w_a$ FOM} \\
\cline{2-19}
& ww & gg & 3x & ww & gg & 3x & ww & gg & 3x & ww & gg & 3x & ww & gg & 3x & ww & gg & 3x \\
\hline
{\sc ComplexSOM } & 349.5 & 1538.1 & 1540.0 & 25.4 & 1830.7 & 7919.6 & 0.7 & 16.2 & 101.6 & 348.1 & 1556.5 & 1558.1 & 21.5 & 1483.8 & 6603.1 & 0.6 & 13.9 & 100.3\\ 
{\sc JaxCNN } & -- & -- & -- & -- & -- & -- & -- & -- & -- & -- & -- & -- & -- & -- & -- & -- & -- & --\\ 
{\sc JaxResNet } & -- & -- & -- & -- & -- & -- & -- & -- & -- & \textbf{365.5} & 1608.2 & 1609.6 & \textbf{44.7} & 2409.0 & 6438.7 & \textbf{1.1} & 26.3 & 161.5\\ 
{\sc NeuralNetwork1 } & 342.1 & 1513.0 & 1515.6 & 25.0 & 2043.7 & 8666.6 & 0.7 & 18.1 & 109.9 & 340.6 & 1468.6 & 1471.5 & 25.0 & 1793.6 & 7876.3 & 0.7 & 15.9 & 96.9\\ 
{\sc NeuralNetwork2 } & 353.1 & \textbf{1742.6} & \textbf{1744.2} & \textbf{35.8} & 2440.7 & \textbf{10025.7} & 0.9 & 22.4 & 136.1 & 350.4 & \textbf{1841.7} & \textbf{1842.9} & 36.7 & 2882.6 & \textbf{11023.9} & 0.9 & 25.8 & 140.0\\ 
{\sc PCACluster } & * & * & * & * & * & * & * & * & * & * & * & * & * & * & * & * & * & *\\ 
{\sc ZotBin } & 350.0 & 1560.2 & 1562.2 & 29.1 & 2449.1 & 8013.3 & 0.7 & 23.3 & 135.3 & 357.9 & 1740.5 & 1741.8 & 38.1 & 2787.5 & 9546.9 & 0.9 & 26.2 & 154.7\\ 
{\sc ZotNet } & \textbf{354.3} & 1574.0 & 1575.8 & 34.6 & \textbf{2604.4} & 8150.8 & \textbf{0.9} & \textbf{25.0} & \textbf{145.9} & 363.1 & 1732.6 & 1733.9 & 43.0 & \textbf{3278.9} & 9226.3 & 1.0 & \textbf{31.4} & \textbf{167.2}\\ 
\hline
{\sc AutokerasLSTM } & -- & -- & -- & -- & -- & -- & -- & -- & -- & 364.3 & 1659.1 & 1660.7 & 44.6 & 1602.4 & 5864.7 & 1.0 & 15.8 & 98.4\\ 
{\sc CNN } & 354.5 & 1628.9 & 1630.5 & 38.1 & 1808.9 & 7000.4 & 0.9 & 16.9 & 103.3 & 366.0 & 1731.1 & 1732.3 & 48.7 & 2074.6 & 7725.6 & 1.1 & 19.8 & 122.4\\ 
{\sc ENSEMBLE } & * & * & * & * & * & * & * & * & * & * & * & * & * & * & * & * & * & *\\ 
{\sc FunBins } & 355.7 & 1497.2 & 1499.2 & \textbf{39.8} & \textbf{3068.2} & \textbf{10959.1} & \textbf{0.9} & \textbf{25.6} & \textbf{141.8} & 367.9 & 1658.3 & 1659.7 & \textbf{54.0} & \textbf{3751.4} & \textbf{12668.2} & \textbf{1.2} & \textbf{31.4} & \textbf{167.2}\\ 
{\sc GPzBinning } & 358.2 & 1505.4 & 1507.4 & 35.1 & 1825.8 & 7197.6 & 0.9 & 16.9 & 111.7 & 366.5 & 1700.7 & 1701.9 & 47.8 & 2129.8 & 8064.7 & 1.1 & 20.2 & 126.9\\ 
{\sc IBandOnly } & 330.3 & 854.7 & 866.6 & 3.6 & 583.1 & 1829.9 & 0.2 & 6.1 & 57.3 & 330.3 & 854.7 & 866.6 & 3.6 & 583.1 & 1829.9 & 0.2 & 6.1 & 57.3\\ 
{\sc LGBM } & 356.3 & 1638.9 & 1640.5 & 39.6 & 1896.2 & 7215.1 & 0.9 & 17.8 & 108.2 & 365.8 & 1734.4 & 1735.6 & 48.6 & 2122.0 & 7933.4 & 1.1 & 20.2 & 125.2\\ 
{\sc LSTM } & 354.2 & 1629.1 & 1630.8 & 38.2 & 1818.0 & 7008.3 & 0.9 & 17.0 & 103.6 & 366.1 & 1729.3 & 1730.5 & 48.9 & 2125.5 & 7997.3 & 1.1 & 20.3 & 126.9\\ 
{\sc MLPQNA } & 355.1 & 1674.8 & 1676.5 & 38.0 & 2177.1 & 8624.5 & 0.9 & 20.0 & 121.8 & 367.5 & 1809.2 & 1810.4 & 51.4 & 2388.7 & 9076.2 & 1.2 & 22.3 & 133.7\\ 
{\sc Stacked Generalization } & 351.6 & 1670.1 & 1672.2 & 39.1 & 2036.3 & 8038.7 & 0.9 & 19.3 & 120.4 & 358.8 & \textbf{1837.8} & \textbf{1839.4} & 49.3 & 2601.7 & 9386.8 & 1.1 & 25.0 & 148.9\\ 
{\sc PQNLD } & 358.7 & 1583.4 & 1585.1 & 36.4 & 1869.4 & 7666.6 & 0.9 & 16.8 & 105.4 & 368.2 & 1801.3 & 1802.5 & 52.9 & 2419.2 & 9353.9 & 1.2 & 22.4 & 133.8\\ 
{\sc Random } & 323.5 & 646.3 & 670.1 & 0.0 & 0.6 & 17.7 & 0.0 & 0.0 & 1.2 & 323.5 & 646.3 & 670.1 & 0.0 & 0.6 & 17.7 & 0.0 & 0.0 & 1.2\\ 
{\sc RandomForest } & 355.6 & 1687.5 & 1689.2 & 39.6 & 2156.0 & 8447.7 & 0.9 & 19.9 & 118.5 & 368.1 & 1823.9 & 1825.1 & 53.2 & 2448.6 & 9265.4 & 1.2 & 22.9 & 136.5\\ 
{\sc SimpleSOM } & \textbf{358.7} & 1572.8 & 1574.4 & 36.7 & 1785.8 & 7244.5 & 0.9 & 16.0 & 98.5 & \textbf{368.4} & 1788.5 & 1789.7 & 53.8 & 2446.7 & 9279.8 & 1.2 & 22.6 & 134.8\\ 
{\sc FFNN } & 355.2 & 1671.3 & 1673.0 & 37.7 & 2064.0 & 8170.0 & 0.9 & 19.0 & 114.7 & 367.3 & 1798.6 & 1799.8 & 50.9 & 2435.6 & 9332.9 & 1.1 & 22.7 & 137.2\\ 
{\sc TCN } & 355.3 & \textbf{1693.1} & \textbf{1694.7} & 38.7 & 2055.4 & 7955.5 & 0.9 & 18.9 & 110.5 & 366.2 & 1814.3 & 1815.5 & 49.4 & 2401.1 & 9100.2 & 1.1 & 22.5 & 134.0\\ 
{\sc UTOPIA } & 353.2 & 1612.4 & 1614.4 & 36.3 & 2022.5 & 8148.8 & 0.9 & 18.5 & 112.2 & 367.6 & 1807.4 & 1808.6 & 52.3 & 2481.2 & 9537.2 & 1.2 & 23.1 & 139.3\\ 

\hline
\end{tabular}
\caption{The 9-bin results for all metrics calculated for all methods on the CosmoDC2 sample.
\textit{ww} are weak-lensing metrics, \textit{gg} clustering, and \textit{3x} the full 3x2pt.
\label{tab:full_dc2}
}
\end{sidewaystable}

\begin{sidewaystable}
\vspace{10cm}
\begin{tabular}{ |l | c c c | c c c | c c c | c c c | c c c | c c c |}
\hline
\multirow{3}{*}{Method} & \multicolumn{9}{|c|}{riz} & \multicolumn{9}{c|}{griz} \\
\cline{2-19}
&  \multicolumn{3}{c|}{SNR}   &  \multicolumn{3}{c|}{$\sigma_8-\Omega_m$ FOM} & \multicolumn{3}{c|}{$w_0-w_a$ FOM}
&  \multicolumn{3}{c|}{SNR}   &  \multicolumn{3}{c|}{$\sigma_8-\Omega_m$ FOM} & \multicolumn{3}{c|}{$w_0-w_a$ FOM} \\
\cline{2-19}
& ww & gg & 3x & ww & gg & 3x & ww & gg & 3x & ww & gg & 3x & ww & gg & 3x & ww & gg & 3x \\
\hline
{\sc ComplexSOM } & 243.9 & 1659.7 & 1660.1 & 8.4 & 1728.2 & 5372.0 & 0.4 & 16.5 & 74.9 & 256.3 & 1826.0 & 1826.4 & 15.9 & 1462.7 & 5177.9 & 0.6 & 14.3 & 75.0\\ 
{\sc JaxCNN } & -- & -- & -- & -- & -- & -- & -- & -- & -- & 258.6 & 1830.2 & 1830.5 & 15.5 & 2152.7 & 6143.8 & 0.6 & 21.2 & 95.4\\ 
{\sc JaxResNet } & -- & -- & -- & -- & -- & -- & -- & -- & -- & -- & -- & -- & -- & -- & -- & -- & -- & --\\ 
{\sc NeuralNetwork1 } & \textbf{253.2} & 1729.1 & 1729.5 & \textbf{13.3} & \textbf{2168.6} & \textbf{7304.1} & \textbf{0.5} & \textbf{19.9} & \textbf{93.7} & \textbf{260.7} & 1773.2 & 1773.5 & \textbf{17.9} & 2502.8 & \textbf{7405.5} & \textbf{0.6} & 23.4 & \textbf{105.4}\\ 
{\sc NeuralNetwork2 } & 251.5 & \textbf{1810.4} & \textbf{1810.9} & 12.6 & 1705.4 & 6279.2 & 0.5 & 16.0 & 76.5 & 258.9 & \textbf{1948.7} & \textbf{1949.0} & 14.1 & 1417.1 & 4962.9 & 0.5 & 14.0 & 73.9\\ 
{\sc PCACluster } & * & * & * & * & * & * & * & * & * & * & * & * & * & * & * & * & * & *\\ 
{\sc ZotBin } & 247.9 & 1674.3 & 1674.9 & 10.5 & 1584.9 & 5448.1 & 0.4 & 15.1 & 76.5 & 253.7 & 1787.2 & 1787.6 & 12.4 & 2084.0 & 6086.2 & 0.5 & 20.1 & 89.4\\ 
{\sc ZotNet } & 249.7 & 1673.3 & 1673.9 & 11.7 & 1897.9 & 6755.9 & 0.5 & 17.4 & 86.0 & 256.3 & 1727.2 & 1727.6 & 13.4 & \textbf{2672.5} & 6686.8 & 0.5 & \textbf{25.0} & 98.4\\ 
\hline
{\sc AutokerasLSTM } & -- & -- & -- & -- & -- & -- & -- & -- & -- & 255.4 & 1844.6 & 1845.0 & 11.5 & 1499.6 & 5123.5 & 0.5 & 14.4 & 72.2\\ 
{\sc CNN } & 252.2 & 1724.5 & 1725.0 & 13.2 & 1335.4 & 4789.3 & 0.5 & 13.1 & 70.3 & 259.0 & 1814.3 & 1814.6 & 16.2 & 1364.9 & 4641.3 & 0.6 & 13.7 & 74.5\\ 
{\sc ENSEMBLE } & 253.1 & 1761.4 & 1761.9 & \textbf{13.7} & 1470.2 & 5293.3 & \textbf{0.5} & 14.2 & 74.6 & 261.1 & 1901.2 & 1901.5 & 17.9 & 1609.7 & 5443.5 & 0.6 & 15.9 & 82.9\\ 
{\sc FunBins } & 251.1 & 1412.5 & 1413.1 & 11.8 & 1297.3 & 4542.4 & 0.5 & 12.4 & 72.2 & 260.1 & 1588.2 & 1588.6 & 17.4 & 1913.5 & 6017.6 & 0.6 & 18.1 & 91.7\\ 
{\sc GPzBinning } & \textbf{254.1} & 1561.6 & 1562.1 & 12.5 & 1328.6 & 4958.8 & 0.5 & 12.8 & 70.7 & 260.7 & 1788.9 & 1789.2 & 16.7 & 1524.0 & 5261.6 & 0.6 & 15.3 & 82.5\\ 
{\sc IBandOnly } & 222.7 & 755.5 & 763.7 & 0.4 & 227.8 & 867.6 & 0.0 & 2.2 & 22.0 & 222.7 & 755.5 & 763.7 & 0.4 & 227.8 & 867.6 & 0.0 & 2.2 & 22.0\\ 
{\sc LGBM } & 252.5 & 1732.6 & 1733.0 & 13.4 & 1345.8 & 4807.2 & 0.5 & 13.2 & 70.7 & 259.2 & 1814.6 & 1814.9 & 16.5 & 1382.5 & 4658.0 & 0.6 & 13.9 & 75.4\\ 
{\sc LSTM } & 252.3 & 1724.0 & 1724.5 & 13.3 & 1330.7 & 4764.1 & 0.5 & 13.0 & 69.9 & 259.2 & 1815.0 & 1815.3 & 16.5 & 1366.0 & 4620.1 & 0.6 & 13.8 & 74.7\\ 
{\sc MLPQNA } & 252.9 & 1757.1 & 1757.6 & 13.7 & 1479.6 & 5300.1 & 0.5 & 14.3 & 73.2 & 261.3 & \textbf{1917.1} & \textbf{1917.4} & 18.3 & 1659.0 & 5637.5 & 0.6 & 16.4 & 83.4\\ 
{\sc Stacked Generalization } & 252.3 & 1682.0 & 1682.6 & 13.1 & \textbf{1748.2} & \textbf{6796.2} & 0.5 & \textbf{16.3} & \textbf{85.3} & 259.1 & 1779.7 & 1780.1 & 17.6 & \textbf{2254.6} & \textbf{7736.9} & 0.6 & \textbf{21.1} & \textbf{97.2}\\ 
{\sc PQNLD } & 254.1 & 1609.2 & 1609.6 & 12.7 & 1406.6 & 5281.2 & 0.5 & 13.4 & 70.6 & 261.8 & 1858.2 & 1858.5 & 18.3 & 1611.9 & 5593.7 & 0.6 & 15.9 & 84.3\\ 
{\sc Random } & 221.5 & 674.0 & 680.5 & 0.0 & 0.4 & 6.3 & 0.0 & 0.0 & 0.8 & 221.5 & 674.0 & 680.5 & 0.0 & 0.4 & 6.3 & 0.0 & 0.0 & 0.8\\ 
{\sc RandomForest } & 252.1 & 1721.9 & 1722.5 & 13.1 & 1462.0 & 5272.6 & 0.5 & 14.1 & 72.5 & 261.3 & 1913.0 & 1913.3 & 18.2 & 1658.2 & 5647.6 & 0.6 & 16.3 & 83.5\\ 
{\sc SimpleSOM } & 254.0 & 1608.6 & 1609.1 & 12.7 & 1415.2 & 5352.9 & 0.5 & 13.5 & 70.7 & \textbf{261.8} & 1830.6 & 1830.9 & \textbf{18.5} & 1622.2 & 5685.9 & \textbf{0.6} & 15.9 & 84.7\\ 
{\sc FFNN } & 252.6 & 1744.7 & 1745.2 & 13.5 & 1465.8 & 5251.9 & 0.5 & 14.1 & 72.4 & 260.8 & 1910.9 & 1911.2 & 17.7 & 1668.3 & 5677.1 & 0.6 & 16.4 & 83.5\\ 
{\sc TCN } & 252.3 & \textbf{1761.4} & \textbf{1762.0} & 13.4 & 1514.0 & 5414.2 & 0.5 & 14.6 & 74.6 & 259.1 & 1913.4 & 1913.7 & 16.4 & 1651.7 & 5613.3 & 0.6 & 16.3 & 82.4\\ 
{\sc UTOPIA } & 252.1 & 1658.9 & 1659.3 & 12.1 & 1457.9 & 5354.9 & 0.5 & 13.9 & 73.6 & 261.2 & 1865.4 & 1865.7 & 17.2 & 1632.7 & 5637.6 & 0.6 & 16.0 & 83.1\\ 

\hline
\end{tabular}
\caption{The 9-bin results for all metrics calculated for all methods on the Buzzard sample.
\textit{ww} are weak-lensing metrics, \textit{gg} clustering, and \textit{3x} the full 3x2pt.
\label{tab:full_buzz}
}
\end{sidewaystable}

\end{document}